\documentclass[11pt,a4paper]{article}
\usepackage{jheppub}
\usepackage[dvipsnames, svgnames, x11names]{xcolor}
\usepackage{lmodern}
\usepackage[utf8]{inputenc}
\usepackage{color}
\usepackage{float}
\usepackage{amsmath,bm}
\usepackage{amssymb}
\usepackage{graphicx}
\usepackage{setspace}
\usepackage{subfigure}
\usepackage{amsthm}
\usepackage{amsfonts}
\usepackage{braket}
\usepackage{multirow}
\usepackage{ulem}
\usepackage{slashed}
\usepackage{babel}
\usepackage{lipsum}
\usepackage{hyperref}
\usepackage{url}
\hypersetup{
	colorlinks=true,
	linkcolor=cyan,
	urlcolor=blue,
	citecolor=red,
}

\def\bra#1{\left\langle#1\right|}
\def\ket#1{\left|#1\right\rangle}

\def\abs#1{\left|#1\right|}
\def\kc#1{\left(#1\right)}

\def\ke#1{\left\{#1\right\}}

\def\Tr{\mathrm{Tr}}

\def\E{\mathcal{E}}
\def\I{\mathbbm{I}}
\def\H{\mathcal{H}}

\def\B{\mathcal{B}}

\def\I{\mathcal{I}}

\def\R{\mathcal{R}}
\def\S{\mathcal{S}}

\title{Boundary mutual information in double holography}

\author[a]{Yuxuan Liu,}

\author[b,c]{Yi Ling,}

\author[d]{and Zhuo-Yu Xian}

\affiliation[a]{
Institute of Quantum Physics, School of Physics, Central South University, Changsha 410083, China
}

\affiliation[b]{School of Physical Sciences, University of Chinese Academy of Sciences, Beijing 100049, China}

\affiliation[c]{Institute of High Energy Physics, Chinese Academy of Sciences, Beijing 100049, China}

\affiliation[d]{Department of Physics, Freie Universit\"at Berlin, Arnimallee 14, DE-14195 Berlin, Germany}

\emailAdd{liuyuxuan93@csu.edu.cn}
\emailAdd{lingy@ihep.ac.cn}
\emailAdd{zhuo-yu.xian@fu-berlin.de}

\abstract{
We consider a composite system where AdS$_3$ gravity is coupled to a flat heat bath and investigate the mutual information between two subregions on the intersection of the AdS$_3$ and bath, referred to as the boundary mutual information (BMI).
The corresponding entanglement entropy is captured via quantum extremal surfaces (QES), which holographically be computed by a surface optimization algorithm based on ``Surface Evolver''. 
We focus on both connected and disconnected configurations of the quantum entanglement wedge (Q-EW) in the AdS$_3$ bulk and analyze the finite corrections to the BMI.
Our numerical results reveal a phase transition of the BMI as the separation between two subregions increases. Furthermore, we find that the BMI can naturally be decomposed into two distinct components: a geometric term arising from the areas of the quantum extremal surfaces, and a correction term resulting from bulk quantum fields within the Q-EW. Interestingly, the geometric contribution always exceeds the total BMI, indicating a negative correction from the bulk matter fields. 
This negativity can be understood as the result of subtracting a greater contribution from quantum fields in the connected Q-EW than in the disconnected one. 
We also reproduce the negative contribution of bulk quantum fields to BMI within a random tensor network (RTN) toy model of double holography. Modeling the bulk as a highly mixed state entangled with a large bath leads to a volume-law bulk entropy. In the large bond-dimension limit, the geometric part of the BMI remains non-negative, while the bulk entropy contribution becomes non-positive when the Q-EWs merge.}

\begin{document}

\maketitle

%------------------------------------
%------------------------------------
\section{Introduction}
%------------------------------------
%------------------------------------
Entanglement entropy plays a vital role in quantum information theory and quantum many-body systems, as it quantifies the degree of quantum entanglement in bipartite pure states. The comprehensive exploration is essential for understanding the underlying structure and the dynamics of quantum systems.

For a region $\bm{A}$, entangled with its complement, the entanglement entropy is given by the von Neumann entropy, $S_{\bm{A}} = -\text{Tr}(\rho_{\bm{A}} \log \rho_{\bm{A}})$, where $\rho_{\bm{A}}$ denotes the reduced density matrix obtained by tracing out the degrees of freedom external to $\bm{A}$. Within the framework of the AdS/CFT correspondence \cite{Maldacena:1997re,Gubser:1998bc,Witten:1998qj}, classical spacetime geometry in the bulk is dual to a quantum state defined on the conformal boundary. The Ryu-Takayanagi (RT) formula \cite{Ryu:2006bv,Ryu:2006ef}, and its covariant generalization \cite{Hubeny:2007xt}, provide a geometric realization of entanglement entropy in this context: the entanglement entropy of a region $\bm{A}$ on the boundary is proportional to the area of a minimal (or extremal) surface in the bulk that is homologous to $\bm{A}$ and anchored on the boundary $\partial\bm{A}$.
Moreover, once the quantum fields in the bulk are considered, the RT surface should be generalized to a QES $\gamma_{\bm{A}}$. The holographic entanglement entropy (HEE) of $\bm{A}$ is calculated by extremizing the generalized gravitational entropy functional \cite{Lewkowycz:2013nqa,Engelhardt:2014gca,Witten:2021unn}. 

Recent breakthroughs in the black hole information paradox \cite{Almheiri:2019hni,Almheiri:2019psf,Almheiri:2019yqk,Penington:2019kki,Almheiri:2019qdq,Almheiri:2019psy} have sparked growing interest in composite quantum systems comprising a $(d-1)$-dimensional quantum system $\partial\mathcal{B}$ coupled to a $d$-dimensional heat bath $\partial$ \cite{Liu:2024cmv,Ling:2021vxe,Izumi:2022opi,Suzuki:2022xwv}. This framework is commonly referred to as the \textbf{boundary perspective}.  In its holographic dual, the $(d-1)$-dimensional system $\partial\mathcal{B}$ corresponds to a (quantum) gravity theory in a $d$-dimensional asymptotically Anti-de Sitter (AAdS) spacetime $\mathcal{B}$ — a viewpoint often termed as the \textbf{brane perspective}. 
A key development in this context is applying the QES $\gamma_{\partial\mathcal{B}}$
to describe the entanglement entropy of this $(d-1)$-dimensional system $\partial\mathcal{B}$ \cite{Almheiri:2019hni,Almheiri:2019psf,Almheiri:2019yqk,Penington:2019kki,Almheiri:2019qdq,Almheiri:2019psy}. In this setting, quantum fields propagating in both the gravity and bath regions are typically described by the same conformal field theory (CFT), and their entanglement entropy are computed using the \textbf{island formula} \cite{Almheiri:2019hni,Chen:2019uhq}, which prescribes extremizing the generalized entropy over all candidate island regions and selecting the configuration that yields the minimal value.
Direct computation of the entanglement entropy for quantum fields on curved spacetimes is often technically challenging \cite{Almheiri:2019yqk}. However, when the quantum fields are described by a CFT with a large central charge, the framework of doubly holographic duality becomes applicable. This duality further maps the $d$-dimensional AAdS spacetime $\mathcal{B}$ into a Planck brane embedded in a $(d+1)$-dimensional classical bulk spacetime. The approach is commonly referred to as the \textbf{bulk gravity perspective}. In this scenario, the QES $\gamma_{\partial\mathcal{B}}$ reduces to the standard RT surface $\Gamma_{\partial\mathcal{B}}$, thereby enabling a purely geometric description of the entanglement entropy \cite{Chen:2020uac,Chen:2020hmv,Hernandez:2020nem,Grimaldi:2022suv,Erdmenger:2015spo}. \footnote{See related discussions in \cite{Takayanagi:2011zk,Chu:2018ntx,Miao:2018qkc,Ling:2020laa,Geng:2020fxl,Geng:2020qvw,Krishnan:2020fer,Miao:2020oey,Akal:2020wfl,Akal:2020twv,Omidi:2021opl,Rozali:2019day,Karlsson:2020uga,Balasubramanian:2020xqf,Balasubramanian:2021wgd,Balasubramanian:2020coy,Miyata:2021ncm,Miyata:2021qsm,Marolf:2020rpm,Marolf:2020xie,Balasubramanian:2020jhl,Peng:2021vhs,Peng:2022pfa,Alishahiha:2020qza,Hashimoto:2020cas,Anegawa:2020ezn,Hartman:2020swn,Chen:2020jvn,Bhattacharya:2020uun,Deng:2020ent,Wang:2021woy,He:2021mst,Gautason:2020tmk,Krishnan:2020oun,Sybesma:2020fxg,Chou:2021boq,Chou:2023adi,Hollowood:2021lsw,Suzuki:2022xwv,Suzuki:2022yru,Bhattacharya:2021nqj,Bhattacharya:2021dnd,Caceres:2021fuw,Bhattacharya:2021jrn,Caceres:2020jcn,Chen:2019iro,Balasubramanian:2020hfs,Almheiri:2020cfm,Li:2020ceg,KumarBasak:2020ams,Anderson:2020vwi,Vardhan:2021mdy,Kawabata:2021vyo,Kawabata:2021hac,Geng:2021iyq,Geng:2021mic,Akal:2021dqt,Renner:2021qbe,Engelhardt:2022qts,Afrasiar:2022ebi,Jeong:2023lkc,Ahn:2021chg,Uhlemann:2021nhu,Karch:2022rvr,Erdmenger:2014xya}.}

To investigate the entanglement structure between the $(d-1)$-dimensional system $\partial\mathcal{B}$ and its external environment, or equivalently the entanglement contribution from the quantum fields propagating in the curved spacetime background $\mathcal{B}$, the RT surface corresponding to $(d-1)$-dimensional regions $\bm{A}\subset\partial\mathcal{B}$ must be constructed. However, for $d\geq 3$, this task becomes significantly more challenging due to the complexity of identifying extremal surfaces near the Planck brane. 

For a simple connected region $\bm{A}$, the corresponding RT surface can be numerically constructed as a two-dimensional non-uniform extremal surface, 
by solving a system of partial differential equations in appropriately chosen coordinates \cite{Liu:2023ggg}. 
This method offers a concrete characterization of the entanglement entropy associated with a single region $\bm{A}$ on $\partial\mathcal{B}$. 
In the semi-classical limit, the entanglement entropy of the boundary region $\bm{A}$ in $d=3$ is found to contain a leading contribution with linear divergence and a subleading term with logarithmic divergence. 
Remarkably, this expression corresponds to the area of the classical extremal surface $\Gamma_{\bm{A}}$ in the bulk and admits two physically equivalent interpretations.
From the brane perspective, the leading divergence in the entanglement entropy arises from the linear-law contribution associated with the entanglement between the brane CFT and the bath CFT, while the subleading divergence encodes the geometric contribution from the QES $\gamma_{\bm{A}}$ localized on the brane.
From the boundary perspective, the leading divergence reflects the volume-law entanglement between $\bm{A}$ and the bath $\partial$, whereas the subleading divergence arises from the entanglement between $\bm{A}$ and the rest of the boundary $\partial\mathcal{B}$. 

For multiple disconnected regions, an analogous formula has not yet been established, despite its importance for studying quantum information measures such as $n$-partite information. 
In this work, we begin with the simplest case — BMI.\footnote{{Though our present study focuses on the bipartite configuration as the simplest illustrative example, the Surface Evolver approach used in this paper can be straightforwardly applied to higher multipartite configurations without any limitations.}} Our goals are twofold: first, to analyze how BMI varies with the size of subregions; and second, to investigate the impact of quantum field entanglement on the brane on the holographic BMI.
In this scenario, however, the aforementioned method becomes inadequate due to the difficulty in identifying a coordinate system capable of capturing disjoint configurations. To further investigate this mixed state entanglement of quantum fields on the brane, we adopt a shape optimization method using the Surface Evolver \cite{brakke1992surface,evolverlink}, a numerical tool widely used in material science for minimizing surface energy via gradient descent techniques. This method constructs minimal surfaces by iteratively evolving an initial surface toward its area-minimizing configuration. It has been successfully applied in holographic settings \cite{Fonda:2014cca,Fonda:2015nma,Seminara:2017hhh,Seminara:2018pmr,Cavini:2019wyb}, making it well-suited for analyzing entanglement structures.

The remainder of this paper is organized as follows. In Section~\ref{sec:2}, we introduce the doubly holographic setup. 
We begin by presenting three equivalent perspectives for the gravitational background with a single Planck brane. 
Subsequently, we describe three corresponding perspectives for entanglement entropy and BMI. 
Finally, we examine the universality of holographic entanglement entropy and BMI within the framework of double holography, particularly in the semiclassical gravity limit.

In Section~\ref{sec:3}, we perform
a numerical analysis on the holographic entanglement entropy and BMI for various configurations. 
We also investigate the correction to BMI arising from quantum fields on the brane.

In Section~\ref{sec:RTN}, we briefly review the framework of RTNs and present a RTN-based interpretation of the correction term in the BMI.

Section~\ref{sec:4} concludes the paper with a summary of our findings and a discussion of potential future directions.

%------------------------------------
%------------------------------------
\section{The setup in double holography}\label{sec:2}
%------------------------------------
%------------------------------------
Consider a general $(d+1)$-dimensional asymptotic AdS spacetime truncated by a $d$-dimensional Planck brane $\mathcal{B}$. This brane intersects with the conformal boundary $\partial$ at infinity, forming a $(d-1)$-dimensional junction $\partial\mathcal{B}$ \cite{Almheiri:2019hni,Chen:2020uac,Almheiri:2019yqk,Takayanagi:2011zk}. This setup admits three equivalent descriptions within the framework of double holography.

%------------------------------------
%------------------------------------
\subsection{The gravitational background}
%------------------------------------
%------------------------------------
Within the doubly holographic framework \cite{Almheiri:2019hni,Chen:2020uac}, the first description, referred to as the bulk gravity perspective, is governed by the $(d+1)$-dimensional gravitational action:
\begin{align}\label{eq:Action}
I=&\frac{1}{16\pi G_N^{(d+1)}} \Bigg[ \int d^{d+1}x
\sqrt{-g}\left(R+\frac{d(d-1)}{L^2}\right)+2\int_{\partial}d^{d}x\sqrt{-h_{\partial}}K_{\partial}\nonumber\\
&+ \int_{\mathcal{B}}d^{d}x\sqrt{-h}\left(K-\alpha\right)-\int_{\partial\mathcal{B}}
d^{d-1}x \sqrt{-\Sigma}\, \theta_{0}\Bigg],
\end{align}
where $h_{\partial}$, $K_{\partial}$ denote the induced metric and the extrinsic curvature on the conformal boundary $\partial$. $h$, $K$ correspond to the induced metric and the extrinsic curvature on the Planck brane $\mathcal{B}$. $\alpha$ characterizes the brane tension.
The final term is the junction condition at the intersection $\partial\mathcal{B}$, with $\theta_{0}$ specifying the dihedral angle between the brane $\mathcal{B}$ and the conformal boundary $\partial$, and $\Sigma$ representing the induced metric on the intersection $\partial\mathcal{B}$. 
The tension on the brane can be further modified by introducing a Dvali-Gabadadze-Porrati term \cite{Randall:1999vf,Dvali:2000hr}, enabling controlled tuning of the Newton constant ratio between the brane and the bulk \cite{Chen:2020uac, Ling:2020laa, Liu:2022pan, Liu:2023ggg}.

This bulk gravitational description admits two equivalent dual interpretations:
\begin{description}
    \item [Brane perspective] The $(d+1)$-dimensional gravity is dual to semi-classical gravity on the brane $\mathcal{B}$, coupled to CFTs living on both the brane $\mathcal{B}$ and the heat bath $\partial$ \cite{Gubser:1999vj,Almheiri:2019hni}.
    \item [Boundary perspective]  The combined gravity-plus-bath theory is further dual to lower-dimensional-one quantum system $\partial\mathcal{B}$ coupled to a heat bath $\partial$ \cite{Takayanagi:2011zk,Almheiri:2019hni,Chen:2020uac}.
\end{description}
%------------------------------------
\begin{figure}
  \centering
  \includegraphics[height=0.4\linewidth]{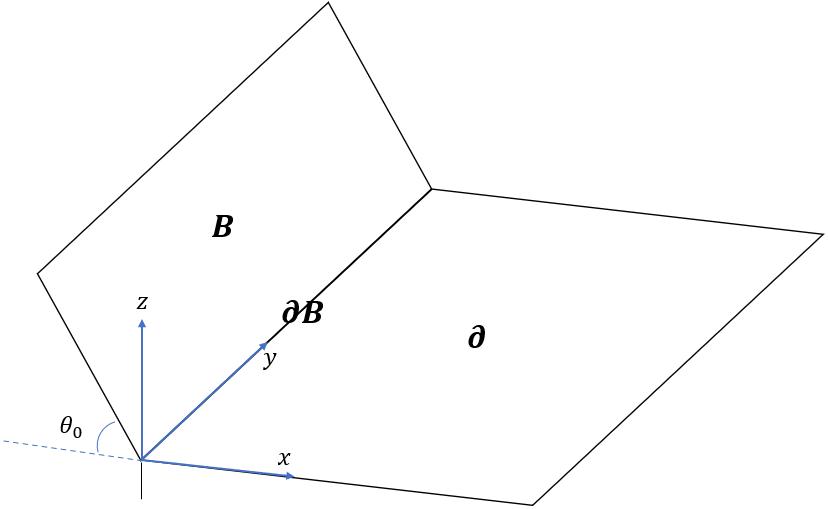}
\caption{The time slice of the AdS$_4$ spacetime with two separated branes.}\label{fig:AdSBrane}
\end{figure}
%------------------------------------
To realize dynamical gravity on the brane, we impose Neumann boundary conditions on the Planck brane $\mathcal{B}$ (see also \cite{Miao:2017gyt,Chu:2017aab,Miao:2018qkc} for alternative viewpoints):
\begin{equation}\label{eq:BCSonBrane_General}
  K_{a b}-K h_{ab}+\alpha h_{ab}=0,
\end{equation}
where $h_{ab}$ is the induced metric on the brane $\mathcal{B}$.

For simplicity, we restrict our analysis to pure $\text{AdS}_{4}$ spacetime with a single brane, corresponding to the vacuum state of the quantum fields living on the boundary. In Poincaré coordinates $(z, x, y, t)$, the metric is given by
\begin{align}\label{eq:AdSmetric}
ds^2=L^2\frac{-dt^2+dz^2+dx^2+d y^2}{z^2},
\end{align}
where $L$ is the AdS radius. The brane is translationally invariant along the $y$-direction -- Fig.~\ref{fig:AdSBrane}, with the position at
\begin{equation}
    x \tan \theta_{0} + z = 0.
\end{equation}
Substitute this constraint into \eqref{eq:AdSmetric}, the induced geometry on the brane is just an AdS$_3$ spacetime, with the metric being given by
\begin{equation}
    ds^2=L^2\frac{-dt^2+dz^2+d y^2}{z^2},
\end{equation}
where $L\to L \sin\theta_0$, $t\to t/\sin\theta_0$ and $y \to y/\sin\theta_0$.
The Neumann boundary condition determining the brane tension (\ref{eq:BCSonBrane_General}) takes the following explicit form in AdS spacetime\footnote{Alternative boundary conditions have been discussed in \cite{Seminara:2017hhh}.}:
\begin{equation}\label{eq:BCSonBrane_AdS}
    \alpha = \frac{2}{L}\cos \theta_{0},
\end{equation}

With this geometry fully specified, we proceed in subsequent sections to construct the RT surfaces corresponding to boundary regions, which encode the BMI.
%------------------------------------
%------------------------------------
\subsection{The entanglement entropy and BMI}\label{sec:EEMI}
%------------------------------------
%------------------------------------

%------------------------------------
\begin{figure}
  \centering
  \subfigure[]{\label{fig:bcft}
  \includegraphics[width=0.45\linewidth]{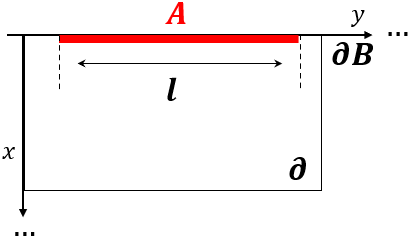}}
    \hspace{0pt}
  \subfigure[]{\label{fig:branecft}
  \includegraphics[width=0.45\linewidth]{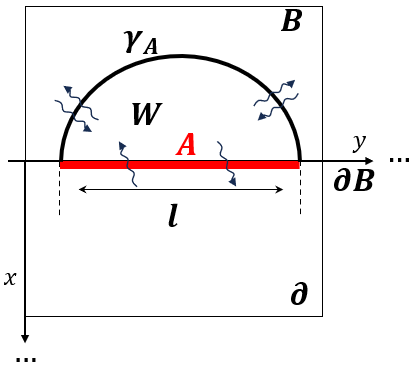}}\\
    \subfigure[]{\label{fig:adscft}
  \includegraphics[width=0.45\linewidth]{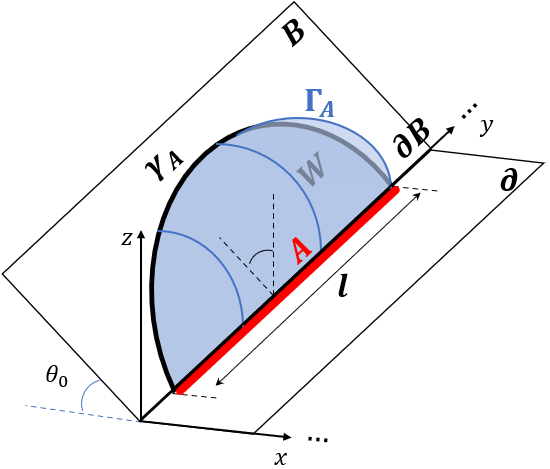}}
\caption{(a) Boundary perspective: $\bm{A}$ is a region on $\partial\mathcal{B}$, located at the boundary of the heat bath $\partial$. (b) Brane perspective: $\gamma_{\bm{A}}$ and $\mathcal{W}$ represent the QES and the corresponding Q-EW on the brane. Curved arrows represent the contributions from quantum fields within $\mathcal{W}$. (c) Bulk gravity perspective: $\Gamma_{\bm{A}}$ represents the classical extremal surface bounded by $\gamma_{\bm{A}} \cup \bm{A}$; $l$ denotes the length of $\bm{A}$.}\label{fig:regionA}
\end{figure}
%------------------------------------
From the boundary perspective, let us consider a $1$-dimensional spatial region $\bm{A}\subset\partial\mathcal{B}$ within a $(2+1)$-dimensional quantum system $\partial\mathcal{B}\cup\bm \partial$ as shown in Fig.~\ref{fig:bcft}. Its entanglement entropy admits two equivalent perspectives within the holographic framework:
\begin{enumerate}
    \item  \textit{Brane perspective} -- Fig.~\ref{fig:branecft}: The entanglement entropy of $\bm{A}$ is holographically given by the island formula
    \begin{equation}\label{eq:EEfromBrane}
        \mathcal{S}[\bm{A}]= \min_{\gamma_{\bm{A}}}\left\{ \frac{\textbf{Area}(\gamma_{\bm{A}})}{4G_{\textbf{eff}}^{(3)}} + S_{\text{QFT}}(\mathcal{W}_{\bm{A}})\right\},
    \end{equation}
    where $\gamma_{\bm{A}}$ denotes the quantum extremal surface on the AdS$_3$ background. The first term on the r.h.s. $$\mathcal{S}_g[\bm{A}]:=\frac{\textbf{Area}(\gamma_{\bm{A}})}{4G_{\textbf{eff}}^{(3)}}$$ captures the geometric contribution, while the second term $$\mathcal{S}_c[\bm{A}]:=S_{\text{QFT}}(\mathcal{W}_{\bm{A}})$$ accounts for the entanglement of the quantum fields within the Q-EW $\mathcal{W}_{\bm{A}}$. Here, $G_{\textbf{eff}}^{(3)}$ is the effective Newton constant on the brane $\mathcal{B}$. The configuration of the quantum extremal surface $\gamma_{\bm{A}}$ has been fully determined by the extremization procedure, and if multiple candidates $\gamma_{\bm{A}}$ locally minimize the entropy, one should choose the one that achieves the global minimum.
    
    \item  \textit{Bulk gravity perspective} -- Fig.~\ref{fig:adscft}: Both terms in (\ref{eq:EEfromBrane}) admit a second holographic dual description to a classical RT surface $\Gamma_{\bm{A}}$ in the higher-dimensional one bulk, with the formula to be
    \begin{equation}\label{eq:QES}
        \mathcal{S}[\bm{A}]=\min_{\Gamma_{\bm{A}}}\left\{\frac{\textbf{Area}(\Gamma_{\bm{A}})}{4G_N^{(4)}}\right\}.
    \end{equation}
    The right-hand side term corresponds to the area of a classical extremal surface $\Gamma_{\bm{A}}$ that extends into the bulk and the surface is bounded by the union of the region $\bm{A}$ and the QES $\gamma_{\bm{A}}$ on the brane. The entanglement entropy of $\bm{A}$ is then obtained by selecting the surface $\Gamma_{\bm{A}}$ associated with the globally minimal generalized entropy among all admissible QES $\gamma_{\bm{A}}$ configurations.
\end{enumerate}

Now consider a region $\bm{A}=\bm{A_1}\cup\bm{A_2}\subset\partial\mathcal{B}$, composed of two disjoint, simply connected subregions $\bm{A_1}$ and $\bm{A_2}$, each residing on the boundary quantum system $\partial\mathcal{B}$, respectively.  The BMI between $\bm{A_1}$ and $\bm{A_2}$ is defined as
\begin{align}\label{eq:mutualinformation}
    \mathcal{I}[\bm{A_1}:\bm{A_2}] &:= \mathcal{S}[\bm{A_1}] +\mathcal{S}[\bm{A_2}] -\mathcal{S}[\bm{A_1}\cup \bm{A_2}].
\end{align}
This expression also admits two equivalent perspectives:
\begin{enumerate}
    \item \textit{Brane perspective}: The BMI naturally decomposes into geometric contribution and quantum field contribution separately as
    \begin{align}
        \mathcal{I}[\bm{A_1}:\bm{A_2}] =& \sum_{i=1}^{2}\Bigg\{\min_{\gamma_{\bm{A_i}}}\left[\frac{\textbf{Area}(\gamma_{\bm{A_i}})}{4\;G_{\textbf{eff}}^{(3)}}+S_{\text{QFT}}(\mathcal{W}_{\bm{A_i}})\right]\Bigg\} \nonumber\\
        &-\min_{\gamma_{\bm{A_1}\cup\bm{A_2}}}\left[\frac{\textbf{Area}(\gamma_{\bm{A_1}\cup\bm{A_2}})}{4\;G_{\textbf{eff}}^{(3)}}+S_{\text{QFT}}(\mathcal{W}_{\bm{A_1}\cup\bm{A_2}})\right],
    \end{align}
   where the geometric contribution is given by
    \begin{equation}\label{eq:AreaTerm}
        \mathcal{I}_g[\bm{A_1}:\bm{A_2}]:=\frac{\textbf{Area}(\gamma_{\bm{A_1}})}{4\;G_{\textbf{eff}}^{(3)}}+\frac{\textbf{Area}(\gamma_{\bm{A_2}})}{4\;G_{\textbf{eff}}^{(3)}}-\frac{\textbf{Area}(\gamma_{\bm{A_1}\cup\bm{A_2}})}{4\;G_{\textbf{eff}}^{(3)}},
    \end{equation}
    while the quantum field contribution is 
    \begin{equation}\label{eq:CorrectionTerm}
        \mathcal{I}_c[\bm{A_1}:\bm{A_2}]:=S_{\text{QFT}}(\mathcal{W}_{\bm{A_1}})+S_{\text{QFT}}(\mathcal{W}_{\bm{A_2}})-S_{\text{QFT}}(\mathcal{W}_{\bm{A_1}\cup\bm{A_2}}).
    \end{equation}
    Each QES $\gamma(\mathcal{K})$, with $\mathcal{K} \in \{\bm{A_1},\bm{A_2},\bm{A_1}\cup\bm{A_2}\}$ 
    is determined by extremizing the corresponding entire entropy functional. The global minimum among all candidates is selected. Notably, the phase transition of BMI in this setting depends not only on geometric contributions, as in classical gravity, but also on the quantum field contributions.

    \item \textit{Bulk gravity perspective}: The combination of quantum extremal surfaces is further dual to the classical extremal surfaces in the higher-dimensional bulk as
\begin{equation}
    \mathcal{I}[\bm{A_1}:\bm{A_2}]= \sum_{i=1}^2\left\{\min_{\Gamma_{\bm{A_i}}}\left[\frac{\textbf{Area}(\Gamma_{\bm{A_i}})}{4 G_N^{(4)}}\right]\right\} -\min_{\Gamma_{\bm{A_1}\cup\bm{A_2}}}\left[\frac{\textbf{Area}(\Gamma_{\bm{A_1}\cup \bm{A_2}})}{4 G_N^{(4)}}\right],
\end{equation}
where $\Gamma(\mathcal{K})$ denotes the RT surface of the subregion $\mathcal{K} \in \{\bm{A_1},\bm{A_2},\bm{A_1}\cup\bm{A_2}\}$ in the bulk.
\end{enumerate}

In the next subsection, we will further analyze the universal behavior of BMI in the semiclassical limit and summarize all associated formulas proved numerically in Sec.~\ref{sec:3}.

%------------------------------------
%------------------------------------
\subsection{The universality in the semiclassical limit}
%------------------------------------
%------------------------------------
In quantum information theory, BMI is a fundamental quantity for characterizing correlations in quantum systems, including quantum entanglement. Within the double holographic framework, BMI similarly serves to quantify entanglement between two black holes or between two SYK clusters~\cite{Liu:2022pan}. In this section, we systematically study the BMI of codimension-three subregions. We begin by reviewing the semiclassical limit and the entropy formula for a single subregion. We then discuss the universal entanglement behavior and summarize the entropy and BMI formulas for multiple subregions in the doubly holographic setup, which will be obtained through numerical analysis in the following sections.

Before proceeding, it is crucial to note that for any boundary region $\bm{A} \subset \partial\mathcal{B}$, the entanglement entropy undergoes a phase transition as a function of the brane parameter -- in our case, the dihedral angle $\theta_{0}$ (see \cite{Liu:2022pan} for more general cases). In particular, only in the subcritical regime ($\theta_{0} < \theta_c\simeq0.64$) does the entanglement entropy become non-vanishing, making this the only case meaningful for studying the entanglement properties of $\bm{A}$ \cite{Liu:2024cmv}. To assure this condition is met, our analysis mainly focuses on the semiclassical limit $ {\theta_0 \ll 1}$, which naturally lies within this subcritical regime.
In this limit, the induced gravitational action on the brane takes the form~\cite{Chen:2020uac,Chen:2020hmv,Hernandez:2020nem,Grimaldi:2022suv}:
\begin{equation}
  I_{\text{eff}}=I_{b}+I_{\text{reg}},
  \end{equation}
where $I_{b}$ corresponds to the terms of the bulk action (\ref{eq:Action}) on the brane:
\begin{equation}
  I_{b}=
  \frac{-1}{8\pi G_N^{(4)}} \int_{\mathcal{B}}d^{3}x\sqrt{-h}\alpha,
\end{equation}
and $I_{\text{reg}}$ is given by \cite{Emparan:1999pm,Chen:2020uac}
\begin{equation}
  I_{\text{reg}}=\frac{1}{16 \pi G_{N}^{(4)}} \int_{\mathcal{B}}  {d^3 x} \sqrt{-h}\left[\frac{4}{L}+L R_h \right]+\mathcal{O}[R_h]^2.
  \end{equation}
Combining the above terms, the effective gravitational action on the brane can be expressed as
\begin{equation}\label{eq:HCgravity}
  I_{\text{eff}}=\frac{1}{16 \pi G_{\mathrm{eff}}^{(3)}} \int  {d^3 x} \sqrt{-h}\left[\frac{2}{\ell_{\mathrm{eff}}^2}+R_h\right]+\mathcal{O}[R_h]^2,
  \end{equation}
with the effective Newton constant and AdS$_3$ radius scale defined as
\begin{equation}\label{eq:effecitve_newton_constant}
  \frac{1}{G_{\mathrm{eff}}^{(3)}}=\frac{L}{G_{N}^{(4)}},\quad\text{and}\quad \frac{1}{l_\text{eff}^2}=\frac{2-\alpha
   L}{L^2}.
\end{equation}
In the semiclassical limit, the ratio of the central charges admits a bulk geometric interpretation through the relation as
\begin{align}\label{eq:ratio_of_central_charges}
  \frac{c'}{c}& \simeq 6\sqrt{\frac{1}{2-2\cos \theta_{0}}},\nonumber\\
  \text{with}\quad c' &:= \frac{3}{2}\frac{l_{\text{eff}}}{G_{\text{eff}}^{(3)}} \quad\text{and}\quad c :=\frac{L^{2}}{4G_{N}^{(4)}},
\end{align}
where $c'$ denotes the central charge of the boundary CFT on $\bm \partial \mathcal{B}$, and $c$ corresponds to the central charge of the bath CFT on $\partial$. 

For a one-dimensional single region $\bm{A}=\bm{A_i}\subset \partial\mathcal{B}$, the corresponding entropy formula has been confirmed to take the form as \cite{Liu:2023ggg}
\begin{align}\label{eq:entanglemententropy}
  \mathcal{S}[\bm{A_i}]=&c \frac{l_i}{\epsilon}+\frac{c'}{3}\log \frac{l_i}{\delta}+\mathcal{F}_{\bm{A_i}}, \quad \text{with } i=1,2
\end{align}
where $l$ is the length of $\bm{A_i}$, $c$ denotes the central charge of the CFT$_3$ on both the brane $\mathcal{B}$ and bath $\partial$, while $c'$ represents the central charge of the CFT$_2$ on the boundary $\partial\mathcal{B}$. Moreover, $\epsilon$ and $\delta=\csc\theta_0 \epsilon$ are the UV-cutoffs of the CFT$_3$ and CFT$_2$, respectively \cite{Chen:2020uac, Suzuki:2022xwv}.  {Strictly speaking, our analysis concerns a regime with a small but finite $\theta_0$. We assume a cutoff hierarchy $\epsilon \ll \theta_0 \ll 1$, ensuring that the boundary cutoff $\delta \simeq \epsilon/\theta_0$ remains sufficiently small to act as a proper UV cutoff.}
It is important to note that while the entire entropy formula in~\eqref{eq:entanglemententropy} is derived from computing the area of the classical extremal surface $\Gamma_{\bm{A_i}}$, the geometric and quantum components can be isolated in the semiclassical limit as
\begin{equation}
    \mathcal{S}_g[\bm{A_i}] \to \frac{c'}{3}\log \frac{l_i}{\delta}, \quad\text{and}\quad\mathcal{S}_c[\bm{A_i}] \to c \;\frac{l_i}{\epsilon} + \mathcal{F}_{\bm{A_i}}.
\end{equation} 
From the brane perspective, $\mathcal{S}_g[\bm{A_i}]$ arises from the area of the QES $\gamma_{\bm{A_i}}$, encoding the geometric contribution from the induced gravity on the brane, while $\mathcal{S}_c[\bm{A_i}]$ captures additional entanglement from quantum fields within the Q-EW. The dominant contribution in $\mathcal{S}_c[\bm{A_i}]$ arises from the first linear-law divergent term due to the entanglement between the brane CFT$_3$ and the bath CFT$_3$ near the induced AdS$_3$ boundary, where the spacetime is infinitely stretched. Moreover, the term $\mathcal{F}_{\bm{A_i}}$ is the finite, cutoff-independent contribution from the quantum fields deep into the brane within the Q-EW $\mathcal{W}_{\bm{A}}$.

From the boundary perspective,  {the entanglement entropy of $\bm{A_i}$ naturally decomposes into different sectors corresponding to distinct degrees of freedom: $\mathcal{S}_g[\bm{A_i}]$ arises from the boundary degrees of freedom of $\bm{A_i}$ with the remaining $\partial\mathcal{B}$, while $\mathcal{S}_c[\bm{A_i}]$ reflects the entanglement between the internal degrees of freedom of $\bm{A_i}$ and the bath $\partial$ adjacent to $\bm{A_i}$, which yields a volume-law contribution.}

Furthermore, when considering two disjoint intervals $\bm{A_1}$ and $\bm{A_2}$, the entanglement entropy of the union $\bm{A} = \bm{A_1} \cup \bm{A_2}$ exhibits a phase transition depending on their separation as
\begin{align}\label{eq:entropy2}
    \mathcal{S}[\bm{A_1} \cup \bm{A_2}] =\textbf{min}\Bigg\{& c \frac{l_1 + l_2}{\epsilon}+\frac{c'}{3}\left(\log \frac{l_1}{\delta} + \log \frac{l_2}{\delta}\right) + \mathcal{F}_{\bm{A_1}} + \mathcal{F}_{\bm{A_2}}, \nonumber \\
    & c\frac{l_1 + l_2}{\epsilon}+\frac{c'}{3}\left(\log \frac{l_1 + l_2 + a}{\delta} + \log \frac{a}{\delta}\right) + \mathcal{F}_{\bm{A_1} \cup \bm{A_2}} \Bigg\},
\end{align}
where $l_1$, $l_2$, and $a$ are the lengths of two subregions and their separation, respectively. The terms $\mathcal{F}_\mathcal{K}$, with $\mathcal{K} \in \{\bm{A_1}, \bm{A_2}, \bm{A_1} \cup \bm{A_2}\}$, also represent the finite, cutoff-independent entanglement from the quantum fields inside the wedge $\mathcal{W}_{\bm{A_1}\cup\bm{A_2}}$. Recall that \eqref{eq:entropy2} is examined via the numerical analysis in the next section.
For sufficiently separated subregions (large $a$), the QES of the union $\bm{A_1} \cup \bm{A_2}$  decomposes into two disconnected QES surfaces corresponding to $\bm{A_1}$ and $\bm{A_2}$, respectively, and the entanglement entropy reduces to the sum of individual contributions as described by (\ref{eq:entanglemententropy}). In contrast, when the subregions are adjacent to each other (small $a$), the entropy exhibits a clear deviation from this behavior, signaling a nontrivial phase transition.

Since the UV-divergent part is always addictive as $$\operatorname{Area}(\bm{A_1}) + \operatorname{Area}(\bm{A_2}) = \operatorname{Area}(\bm{A_1}\cup\bm{A_2}),$$ 
the BMI becomes finite and is governed entirely by the finite contributions.
For two nearby subregions, the formula of BMI (\ref{eq:mutualinformation}) reduces to
\begin{align}\label{eq:semiMI}
    \mathcal{I}[\bm{A_1}:\bm{A_2}]=&\frac{c'}{3}\log \frac{l_1 l_2}{(l_1 + l_2 + a) a}+\mathcal{F}_{\bm{A_1} }+\mathcal{F}_{\bm{A_2}} - \mathcal{F}_{\bm{A_1} \cup \bm{A_2}}.
\end{align}
Within this expression, we further have
$$\mathcal{I}_g[\bm{A_1}:\bm{A_2}]=\mathcal{S}_g[\bm{A_1}]+\mathcal{S}_g[\bm{A_2}]-\mathcal{S}_g[\bm{A_1}\cup\bm{A_2}]\to \frac{c'}{3}\log \frac{l_1 l_2}{(l_1 + l_2 + a) a},$$
and
$$\mathcal{I}_c[\bm{A_1}:\bm{A_2}]=\mathcal{S}_c[\bm{A_1}]+\mathcal{S}_c[\bm{A_2}]-\mathcal{S}_c[\bm{A_1}\cup\bm{A_2}]\to\mathcal{F}_{\bm{A_1} }+\mathcal{F}_{\bm{A_2}} - \mathcal{F}_{\bm{A_1} \cup \bm{A_2}}.$$
From the boundary perspective, the first logarithmic term $\mathcal{I}_g[\bm{A_1}:\bm{A_2}]$ denotes the BMI from the CFT$_2$ on the boundary system $\bm \partial \mathcal{B}$, while the second finite term $\mathcal{I}_c[\bm{A_1}:\bm{A_2}]$ represents the correction from the CFT$_3$.
From the brane perspective, the first term $\mathcal{I}_g[\bm{A_1}:\bm{A_2}]$ denotes the geometric contributions from the induced gravity on the brane, which equals the area of the QES,
whereas the second correction term $\mathcal{I}_c[\bm{A_1}:\bm{A_2}]$ therefore encodes the additional cutoff-independent contributions from the CFT$_3$ on the brane.
Interestingly, $\mathcal{I}_c[\bm{A_1}:\bm{A_2}]$ is generally negative from the later numerical analysis. This is mainly because the Q-EW of the union region $\bm{A_1}\cup\bm{A_2}$ is larger than the sum of those associated with $\bm{A_1}$ and $\bm{A_2}$ individually. We will elaborate on this behavior in the next section.

%------------------------------------
\begin{figure}
  \centering
  \subfigure[]{\label{fig:ShapeIni}
  \includegraphics[width=0.45\linewidth]{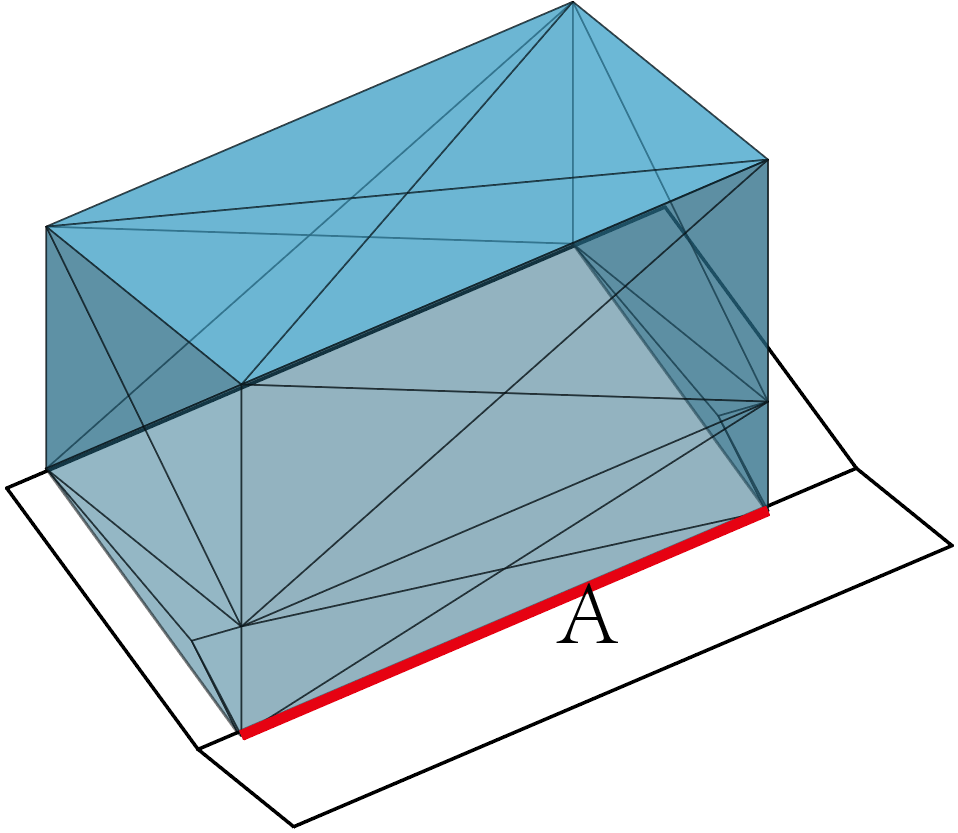}}
    \hspace{0pt}
  \subfigure[]{\label{fig:ShapeFin}
  \includegraphics[width=0.45\linewidth]{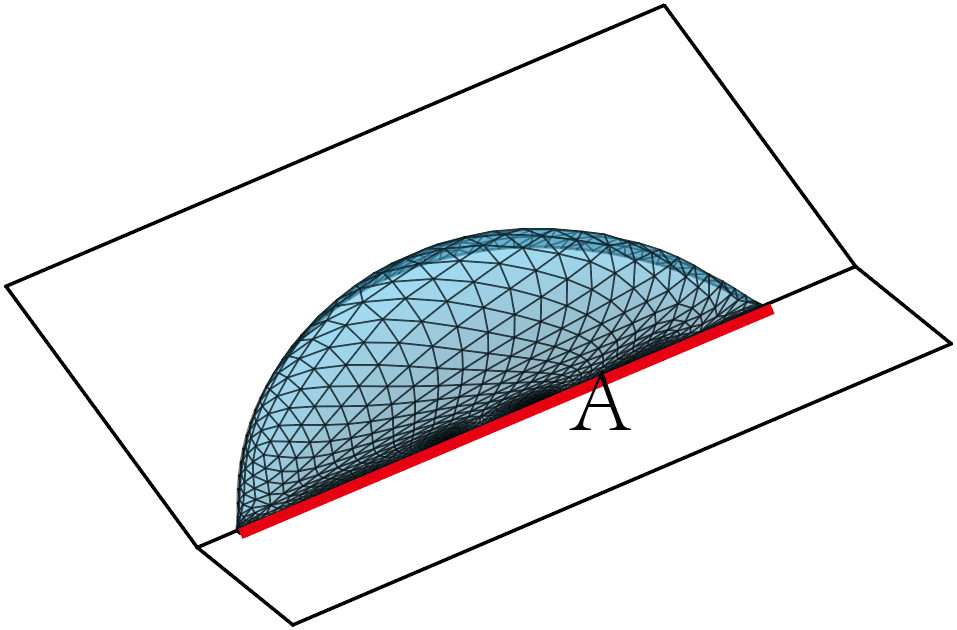}}\\
  \subfigure[]{\label{fig:MIsame3D}
  \includegraphics[height=0.5\linewidth]{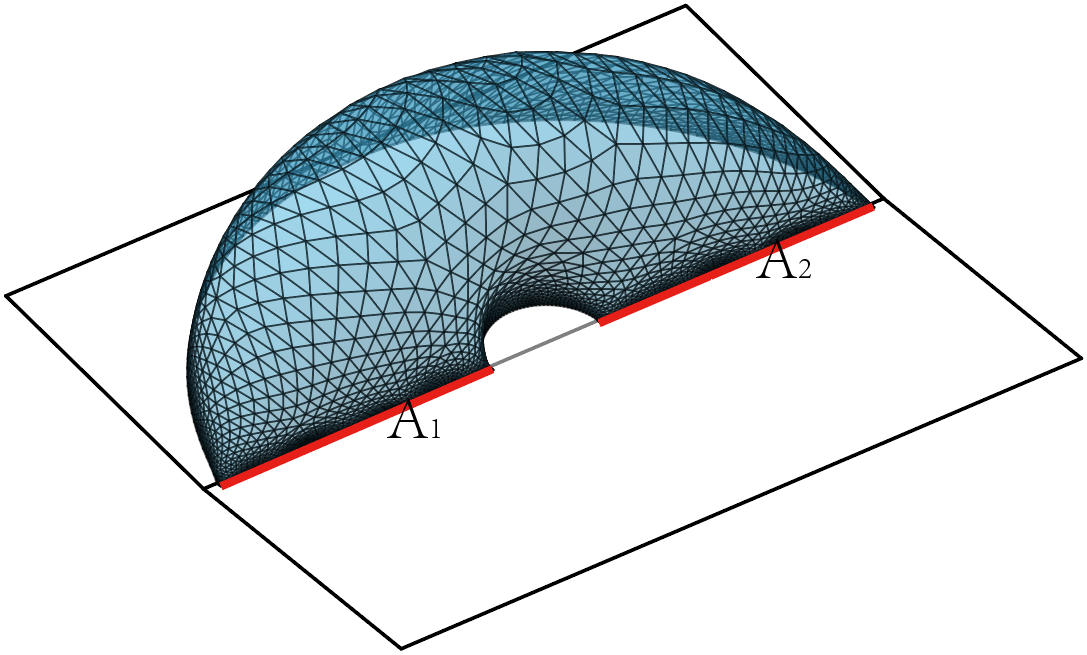}}
\caption{(a): The initial trial surface corresponding to a simply connected boundary region $\bm{A}$, anchored on $z=\epsilon$, constructed as a triangulated mesh. (b): The final minimal surface obtained after the optimization of the triangular facets via gradient descent in Surface Evolver. (c): A connected configuration of the RT surface for a disconnected boundary region $\bm{A}=\bm{A_1}\cup\bm{A_2}$ shown in red, also anchored on $z=\epsilon$. To properly account for all boundary degrees of freedom, the region $\bm{A}$ is regularized as a rectangular strip with nonzero width $x^*\simeq\epsilon\ll \textit{any other length scales}$, following the prescription outlined in \cite{Almheiri:2019hni}.}\label{fig:ShapeEg}
\end{figure}
%------------------------------------(*critical angle!!*)

%------------------------------------
%------------------------------------
\section{Numerical analysis}\label{sec:3}
%------------------------------------
%------------------------------------
In this section, we begin by introducing the engineering software  -- ``Surface Evolver'', which is employed to construct the RT surfaces relevant to our analysis. We then assess the numerical accuracy of the results obtained and verify the emergence of universal entanglement behavior. Finally, we explore the properties of BMI in the semiclassical limit.
%------------------------------------
%------------------------------------
\subsection{Constructions of the RT surfaces}
%------------------------------------
%------------------------------------
Traditionally, constructing the RT surface in the bulk involves solving two-dimensional partial differential equations, often facilitated by adopting suitable coordinate systems to regulate divergences near the boundary region $\bm{A}$ and by imposing Neumann boundary conditions on the brane \cite{Liu:2023ggg}. However, this coordinate-based approach faces inherent limitations when investigating BMI, as it becomes challenging to find appropriate coordinate charts for describing disjoint subregions. To address these challenges, we instead employ the shape optimization software ``Surface Evolver'' \cite{brakke1992surface} to construct the minimal surface $\Gamma_{\bm{A}}$. This method has been successfully applied in various asymptotically AdS spacetimes \cite{Fonda:2014cca,Fonda:2015nma,Seminara:2017hhh,Seminara:2018pmr,Cavini:2019wyb}, offering a flexible and robust numerical framework for studying entanglement structures beyond symmetric or connected regions.

In Surface Evolver, any surface is represented as a collection of oriented triangular facets. Given a background metric, a boundary region $\bm{A}$ is anchored at $z = \epsilon$ to regulate divergences near the asymptotic boundary, and an initial trial surface is constructed -- Fig.\ref{fig:ShapeIni}. The software then evolves this initial configuration toward a local minimum of the area functional via a gradient descent algorithm -- see Fig.~\ref{fig:ShapeFin} (see Appendix~B of \cite{Fonda:2014cca} for an overview of Surface Evolver). The output is a triangulated minimal surface $\Gamma^\epsilon_{\bm{A}}$, anchored at $z = \epsilon$, which approximates the true extremal surface $\Gamma_{\bm{A}}$ (anchored at $z = 0$). The accuracy of this approximation improves with increasing triangulation resolution. For any given triangulated surface, the total area can be directly computed from the mesh data.
For simplicity, in what follows we will not distinguish between the numerically evolved surface $\Gamma^\epsilon_{\bm{A}}$ and the continuum surface $\Gamma_{\bm{A}}$, and will refer to both as $\Gamma_{\bm{A}}$.

Since this work primarily focuses on holographic BMI, we consider configurations where two disjoint subregions $\bm{A_1}$ and $\bm{A_2}$ (with $\bm{A} = \bm{A_1} \cup \bm{A_2}$) are both anchored on the boundary $\bm \partial \mathcal{B}$. In general, the combined region $\bm{A}$ exhibits two distinct entanglement phases:
When the subregions are sufficiently close, the associated Q-EW $\mathcal{W}$ is connected -- Fig.\ref{fig:MIsame3D}. This phase signals strong correlations between $\bm{A_1}$ and $\bm{A_2}$. In contrast, when the subregions are far apart, the RT surface decomposes into a union of the individual RT surfaces corresponding to $\bm{A_1}$ and $\bm{A_2}$, each resembling the configuration shown in Fig.\ref{fig:ShapeFin}, indicating weak correlations.
For visualization purposes, the RT surface shown in Fig.~\ref{fig:MIsame3D} is rendered with a moderate resolution of approximately $V \simeq 3000$ vertices. However, for precise area calculations, we use a higher-resolution discretization with $V \simeq 13000$ vertices to ensure sufficient smoothness and numerical accuracy.

In the subsequent subsections, all entanglement-related quantities will be expressed in dimensionless form by normalizing with respect to the central charge $c=\frac{L^2}{4 G_N^{(4)}}$. Specifically, we define
\begin{align}
    \left\{\mathcal{I}[\bm{A_1}:\bm{A_2}],\mathcal{S}[\mathcal{\mathcal{K}}]\right\}= \frac{1}{c}\left\{ \mathcal{I}[\bm{A_1}:\bm{A_2}],\mathcal{S}[\mathcal{K}]\right\},
\end{align}
where $\mathcal{K} \in \{\bm{A_1},\bm{A_2},\bm{A_1}\cup\bm{A_2}\}$. Accordingly, the geometric and correction terms are also rescaled as
\begin{equation}
    \left\{\mathcal{I}_g,\mathcal{I}_c,\mathcal{S}_g,\mathcal{S}_c\right\}=\frac{1}{c}\left\{\mathcal{I}_g,\mathcal{I}_c,\mathcal{S}_g,\mathcal{S}_c\right\}.
\end{equation}
With these identifications, the effective central charge $c'$ always appears in the dimensionless ratio $c'/c$, which is determined by the dihedral angle via (\ref{eq:ratio_of_central_charges}).

%------------------------------------
%------------------------------------
\subsection{Entanglement entropy of disjoint subregions}
%------------------------------------
%------------------------------------

%------------------------------------
\begin{figure}
  \centering
  \subfigure[]{\label{fig:Sdis}
  \includegraphics[height=0.45\linewidth]{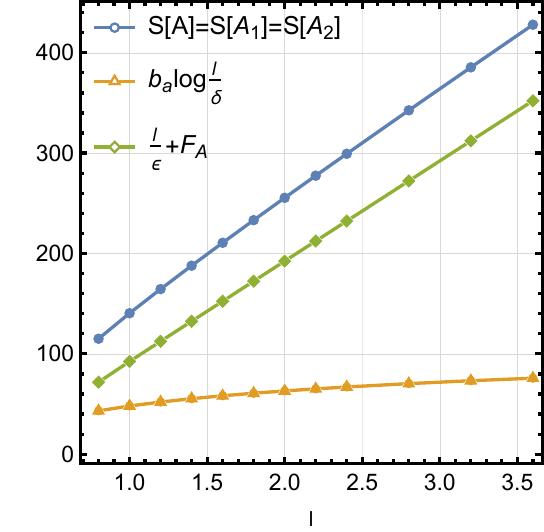}}
    \hspace{0pt}
  \subfigure[]{\label{fig:Cdis}
  \includegraphics[height=0.45\linewidth]{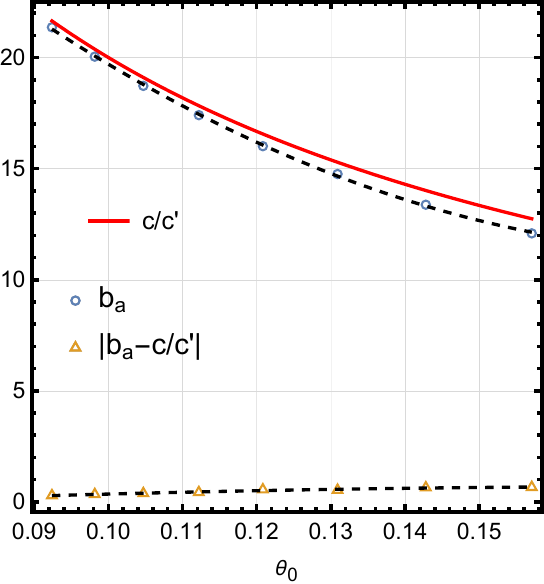}}
\caption{(a): The entanglement entropy of a single region $\bm{A}$ depicted in blue as a function of $l$, where $l$ is the length of the region. The UV cutoff is fixed to be $\epsilon=0.01$, and the dihedral angle is set to $\theta_0=\pi/34$. The fitted geometric and correction contributions are overlaid in yellow and green, respectively. (b): The fitting coefficient $b_a$ associated with the logarithmic divergence is shown in blue, as a function of the dihedral angle $\theta_0$. The red curve represents the theoretical prediction for the ratio of central charges $c'/c$ from (\ref{eq:ratio_of_central_charges}). The difference between the fitting coefficient and the theoretical ratio is illustrated by the yellow dots.}
\end{figure}
%------------------------------------

In this subsection, we consider the entanglement entropy of a disconnected region $\bm{A}=\bm{A_1}\cup\bm{A_2}$. For simplicity,  we restrict our analysis to \textbf{symmetric configurations} with equal subregion lengths $l_1 = l_2 = l$. Throughout, we set the UV cutoff to be $\epsilon=0.01$, which is sufficiently small compared to all other relevant physical scales.

For two \textbf{distant} subregions, the Q-EW associated with the union region $\bm{A_1}\cup\bm{A_2}$ becomes disconnected.
In this regime, the total entanglement entropy is expected to reduce to the sum of the entropies of the individual components.
Thus, we adopt the following fitting function as 
\begin{equation}\label{eq:FitEEDis}
    \frac{1}{2}\mathcal{S}[\bm{A_1}\cup\bm{A_2}]=\mathcal{S}[\bm{A_1}]=\mathcal{S}[\bm{A_2}]=\frac{l}{\epsilon}+b_a \log \frac{l}{\delta}+b_{a1}.
\end{equation}
Here, $b_a$ and $b_{a1}$ are two fitting coefficients. We note that it has been shown that $b_a$ asymptotically approaches the ratio $c'/c$ in the semiclassical limit \cite{Liu:2023ggg}. Therefore, the main aim here is to verify the convergence behavior of our numerical method and assess the robustness of the Surface Evolver-based approach. An illustrative example is presented in Fig.~\ref{fig:Sdis}, where the total entropy (blue curve) is decomposed into a geometric component (yellow) obeying the logarithmic law, and a correction component (green) satisfying the linear law. These numerical behaviors align well with the structure of the fitting function(\ref{eq:FitEEDis}). Moreover, Fig.~\ref{fig:Cdis} demonstrates that the fitting coefficient $b_a$ converges to the theoretical ratio $c'/c$ as $ {\theta_0 \ll 1}$,  thereby confirming the expected semiclassical behavior:
$$\mathcal{S}_g\to b_a \log \frac{l}{\delta},\quad \text{and}\quad \mathcal{S}_c\to \frac{l}{\epsilon}+\mathcal{F}_{\bm{A}}, \quad \text{as}\quad  {\theta_0 \ll 1}.$$

%------------------------------------
\begin{figure}
  \centering
  \subfigure[]{\label{fig:Scon}
  \includegraphics[height=0.45\linewidth]{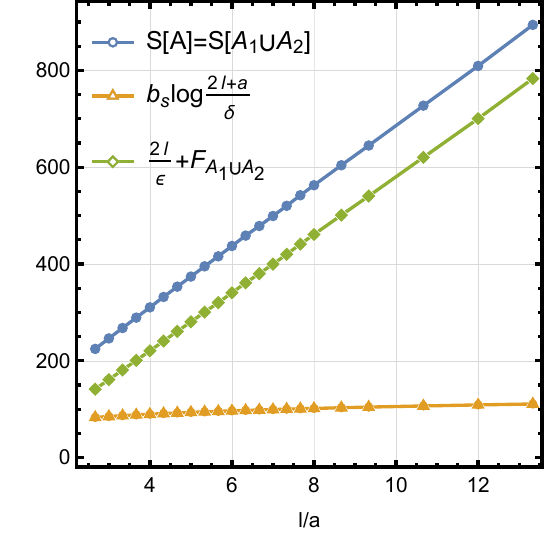}}
    \hspace{0pt}
  \subfigure[]{\label{fig:Ccon}
  \includegraphics[height=0.45\linewidth]{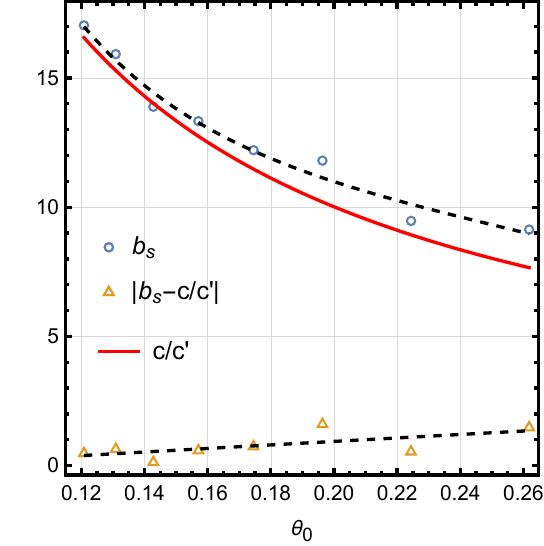}}
\caption{(a): The entanglement entropy of two adjacent subregions $\bm{A}=\bm{A_1} \cup \bm{A_2}$ (blue curve) as a function of the dimensionless ratio $l/a$, with the UV cutoff fixed to be $\epsilon=0.01$, and the dihedral angle $\theta_0=\pi/26$. The numerically extracted geometric contribution (yellow) and correction term (green) are shown separately to illustrate their respective behaviors. (b): The fitting coefficient $b_a$ (blue dots) before the logarithmic divergence term is plotted as a function of the dihedral angle $\theta_0$, while the theoretical ratio of the central charges (red curve) is obtained from (\ref{eq:ratio_of_central_charges}). The discrepancy between the coefficient and the ratio is illustrated by the yellow dots, confirming convergence as $ {\theta_0 \ll 1}$.}
\end{figure}
%------------------------------------

For two \textbf{adjacent} subregions, the Q-EW $\mathcal{W}_{\bm{A_1}\cup\bm{A_2}}$ of the combined region is connected. In this case, we propose the following fitting function for the numerical computation of the entanglement entropy as
\begin{equation}\label{eq:FitEECon}
        \mathcal{S}[\bm{A_1} \cup \bm{A_2}] =
    \frac{2l}{\epsilon} + b_s\left(\log \frac{2l + a}{\delta} + \log \frac{a}{\delta}\right) +b_{s1},
\end{equation}
where $\{b_s,b_{s1}\}$ are fitting coefficients. The linear-divergent term arises from the entanglement of quantum fields near the induced AdS$_3$ boundary. The logarithmic-divergent terms originate from the area of the QES -- Fig.~\ref{fig:MIsame3D}, while other corrections are expressed by the finite term. Within this fitting structure, we identify the geometric and correction contributions as $$\mathcal{S}_g\to b_s\left(\log \frac{2l + a}{\delta} + \log \frac{a}{\delta}\right), \quad \text{and}\quad \mathcal{S}_c\to \frac{2l}{\epsilon}+\mathcal{F}_{\bm{A_1}\cup\bm{A_2}}, \quad \text{as}\quad  {\theta_0 \ll 1}.$$  {As illustrated in Fig.~\ref{fig:Scon},} this decomposition clearly separates the divergent contributions.
Similarly, in the semiclassical limit $ {\theta_0 \ll 1}$, the fitting coefficient $b_s$ converges to the central charge ratio $c'/c$ -- Fig.~\ref{fig:Ccon}. This observation leads to the conclusion that, regardless of whether the Q-EW is connected or disconnected, the coefficient of the logarithmic divergence consistently approaches $c'/c$.

%------------------------------------
\begin{figure}
  \centering
  \subfigure[]{\label{fig:mutualinformation}
  \includegraphics[height=0.45\linewidth]{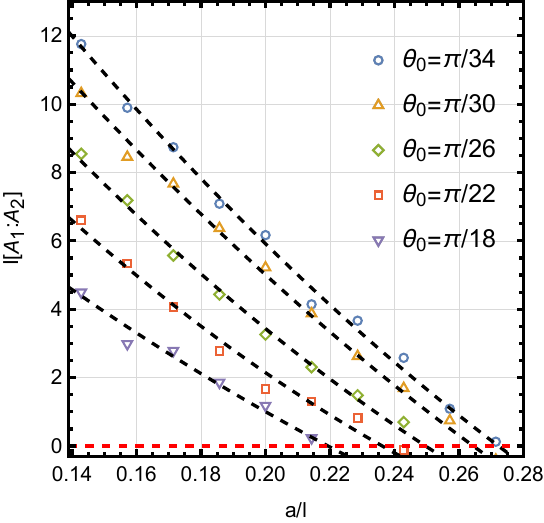}}
    \hspace{0pt}
  \subfigure[]{\label{fig:transitionofMI}
  \includegraphics[height=0.45\linewidth]{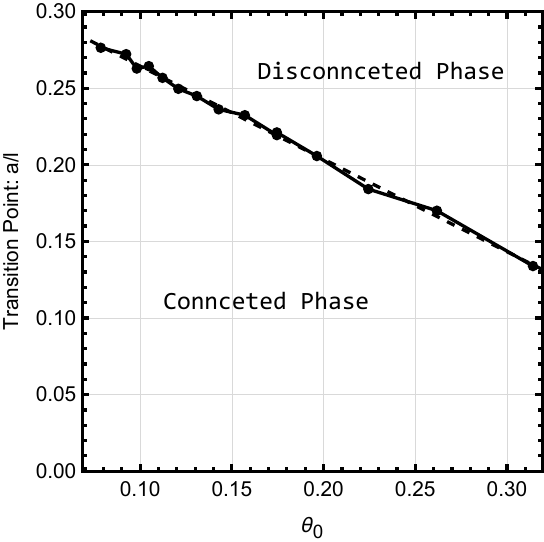}}
\caption{(a): The BMI $\mathcal{I}[\bm{A_1}\cup\bm{A_2}]$ as functions of the dimensionless separation $a/l$. Discrete points of different colors represent the numerical results at different dihedral angles $\theta_0$. The black dashed curves depict the fitting functions of the BMI, and their intersections with the red dashed line mark the transition points of the BMI. (b): The transition points extracted from (a), as a function of the dihedral angle $\theta_0$. Black dots represent numerically determined critical points where the BMI undergoes a phase transition, and the solid black curve interpolates these data points, indicating the transition curve.}\label{fig:MISame}
\end{figure}
%------------------------------------

As a summary, our numerical analysis confirms the expected behavior of the entanglement entropy for disconnected regions in the semiclassical limit. In general, the total entropy is primarily contributed from two distinct sources: First, linear divergence arises from the quantum fields localized near the boundary of the induced gravitational region, reflecting the volume-law entanglement; Second, logarithmic divergence originates from the area of the QES, and corresponds to the entanglement entropy governed by induced gravity brane. 
However, there is an additional finite term that generally captures the contributions from quantum fields deep inside the Q-EW. These contributions are sensitive to the global structure of the QES and are not simply additive, depending on whether the Q-EW is connected or disconnected.
In the following subsection, we will analyze how these finite-term contributions influence the BMI.

%------------------------------------
%------------------------------------
\subsection{BMI of the disjoint subregions}
%------------------------------------
%------------------------------------
In the previous analysis of the entanglement entropy (\ref{eq:entanglemententropy}) and (\ref{eq:entropy2}), our attention was primarily focused on the divergent structures, while the finite contributions were largely neglected due to their sub-subleading nature and the difficulty in obtaining analytical expressions. However, in the case of BMI, all UV divergences are canceled by construction, leaving behind only finite terms. This feature makes BMI a divergent-free quantity, and simultaneously highlights the importance of reliable numerical methods, as these finite terms are generally inaccessible through purely analytical techniques.

Having already computed the relevant entropies, the BMI in the semiclassical limit can now be directly obtained via simple algebraic combinations, as described in (\ref{eq:mutualinformation}). In this section, we proceed to numerically evaluate the BMI between two disjoint subregions $\bm{A_1}$ and $\bm{A_2}$, enabling us to isolate and analyze both the finite geometric contributions and the corrections of quantum fields.

We begin by analyzing the dependence of the BMI $\mathcal{I}[\bm{A_1}:\bm{A_2}]$ on the dimensionless separation $a/l$ between two subregions, as illustrated in Fig.~\ref{fig:mutualinformation}. When the subregions are close, the Q-EW of the union $\bm{A_1} \cup \bm{A_2}$ remains connected, resulting in a non-vanishing (positive) BMI. As the separation increases, the BMI decreases approximately linearly -- a behavior that sharply contrasts with predictions from classical gravity. Eventually, the Q-EW becomes disconnected, and the BMI drops to zero, signaling a phase transition in the underlying geometry.

%------------------------------------
\begin{figure}
  \centering
  \subfigure[]{\label{fig:MIpartition}
  \includegraphics[height=0.45\linewidth]{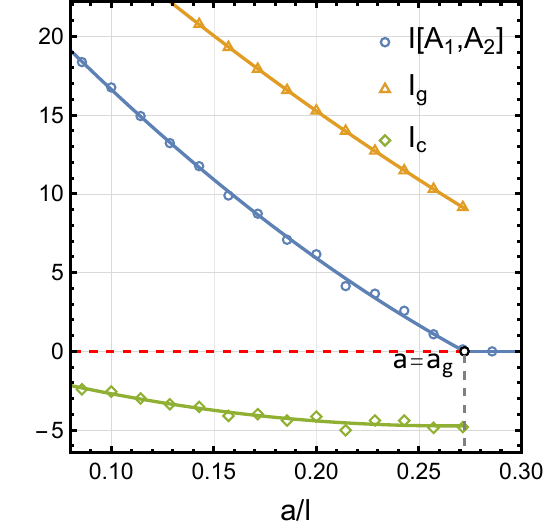}}
    \hspace{0pt}
  \subfigure[]{\label{fig:MIonBraneSame}
  \includegraphics[width=0.45\linewidth]{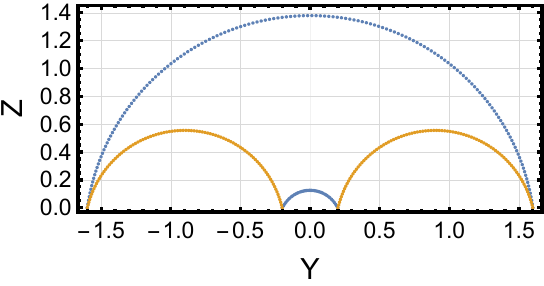}}

\caption{(a): The BMI $\mathcal{I}[\bm{A_1}:\bm{A_2}]=\mathcal{I}_g[\bm{A_1}:\bm{A_2}]+\mathcal{I}_c[\bm{A_1}:\bm{A_2}]$ is plotted in blue as a function of the dimensionless separation $a/l$ between two subregions $\bm{A_1}$ and $\bm{A_2}$, with fixed parameters to be $\{\epsilon,\theta_0\}=\{0.01,\pi/34\}$. The geometric contribution $\mathcal{I}_g[\bm{A_1}:\bm{A_2}]$ is obtained from the direct computation of the area of the corresponding QES on the brane, and is shown in yellow. While the correction $\mathcal{I}_c[\bm{A_1}:\bm{A_2}]=\mathcal{I}[\bm{A_1}:\bm{A_2}]-\mathcal{I}_g[\bm{A_1}:\bm{A_2}]$ is represented by the green curve. (b): The sketch of two candidate QESs for $\{a,l,\epsilon,\theta_0\}=\{0.4,1.4,0.01,\pi/34\}$. 
% The coordinates on the brane can be obtained by $\{Z,Y\}=\{(z-\epsilon)/\sin\theta_0,y\}$. 
As required by the Surface Evolver implementation, all extremal surfaces are anchored on the constant-$z$ slice at $z = \epsilon$.}
\end{figure}
%------------------------------------

Intriguingly, the transition from the connected phase to disconnected phase
occurs at a larger separation $a/l$ when the dihedral angle $\theta_0$ decreases -- Fig.~\ref{fig:transitionofMI}. On one hand, the region above the black curve corresponds to configurations where the Q-EW of $\bm{A_1} \cup \bm{A_2}$ remains connected, indicating strong entanglement between these two subregions. On the other hand, below the curve, Q-EW becomes disconnected, signaling that the entanglement between $\bm{A_1}$ and $\bm{A_2}$ is weak.
This trend suggests that in the semiclassical limit, the entanglement between two subregions is stronger and can persist over larger separations. This observation aligns with the interpretation that decreasing $\theta_0$ corresponds to an increase in the degrees of freedom (d.o.f.) on the (1+1)-dimensional boundary quantum system $\partial\mathcal{B}$ \cite{Ling:2020laa, Liu:2022pan, Liu:2023ggg}. As the ratio $c'/c$ increases, each subregion, such as $\bm{A_1}$, possesses more internal d.o.f. to become entangled with $\bm{A_2}$, thereby extending the range of significant correlations and shifting the transition point outward.

Furthermore, as shown in Fig.~\ref{fig:MIpartition}, the geometric contribution $\mathcal{I}_g[\bm{A_1}:\bm{A_2}]$ decreases monotonically with the increase of the separation. Remarkably, this geometric contribution consistently exceeds the total BMI $\mathcal{I}[\bm{A_1}:\bm{A_2}]$, implying that the correction term is always non-positive:
\begin{align*}
    \mathcal{I}_c[\bm{A_1}:\bm{A_2}]=\mathcal{F}_{\bm{A_1}}+\mathcal{F}_{\bm{A_2}}-\mathcal{F}_{\bm{A_1}\cup\bm{A_2}}\leq0.
\end{align*}

At first glance, this phenomenon may appear counterintuitive, but it admits a natural interpretation from the brane perspective. Recall that each finite term $\mathcal{F}_{\mathcal{K}}$ represents the cutoff-independent entanglement between the quantum fields contained within the Q-EW $\mathcal{W}_{\mathcal{K}}$ and the environment. Therefore, these entanglements are naturally expected to be proportional to the area of the corresponding wedge $\mathcal{W}_{\mathcal{K}}$ ($\mathcal{K}=\bm{A_1}, \bm{A_2}, \bm{A_1}\cup\bm{A_2}$). As illustrated in Fig.~\ref{fig:MIonBraneSame}, when the wedge $\mathcal{W}_{\bm{A_1}\cup\bm{A_2}}$ is connected, its area always exceeds that of the union of two disconnected wedges $\mathcal{W}_{\bm{A_1}}\cup\mathcal{W}_{\bm{A_2}}$. Consequently, the connected wedge $\mathcal{W}_{\bm{A_1}\cup\bm{A_2}}$ contains a greater number of quantum fields. This difference directly leads to a negative value for the correction term in BMI as 
\begin{align}\label{eq:MI_QFT}
\mathcal{I}_c[\bm{A_1}:\bm{A_2}]=S_{\text{QFT}}(\mathcal{W}_{\bm{A_1}})+S_{\text{QFT}}(\mathcal{W}_{\bm{A_2}})-S_{\text{QFT}}(\mathcal{W}_{\bm{A_1}\cup\bm{A_2}}).
\end{align}
Furthermore, as the separation between these two subregions increases while their sizes remain fixed (i.e., as $a/l$ grows), the area of the connected wedge $\mathcal{W}_{\bm{A_1} \cup \bm{A_2}}$ grows increasingly larger relative to the total area of the disconnected wedges $\mathcal{W}_{\bm{A_1}}\cup\mathcal{W}_{\bm{A_2}}$.
This growing amplifies the negativity of $\mathcal{I}_c[\bm{A_1}:\bm{A_2}]$, causing the correction term to decrease monotonically with $a/l$.

\begin{figure}
  \centering
  \subfigure[]{\label{fig:classical}
  \includegraphics[width=0.4\linewidth]{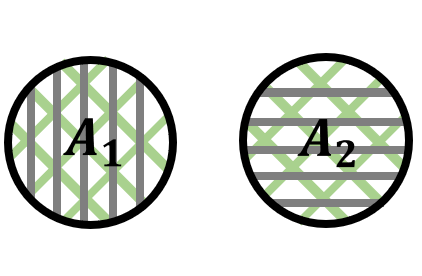}}
    \hspace{50pt}
  \subfigure[]{\label{fig:semiclassical}
  \includegraphics[width=0.4\linewidth]{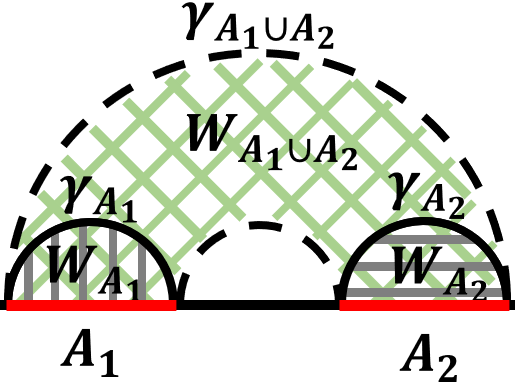}}

\caption{A schematic illustration of the BMI between two subregions $\bm{A_1}$ and $\bm{A_2}$ in the frameworks of (a) standard CFT and (b) double holography, respectively. 
The quantum fields involved in computing the entanglement entropy of $\bm{A_1}$ and $\bm{A_2}$ in (a), and Q-EWs $\mathcal{W}_{\bm{A_1}}$ and $\mathcal{W}_{\bm{A_2}}$ in (b) are indicated by vertical and horizontal gray lines, respectively. The green crossed lines depict the quantum fields involved in computing the entanglement entropy of the union region $\bm{A_1}\cup\bm{A_2}$ in (a), and $\mathcal{W}_{\bm{A_1}\cup\bm{A_2}}$ in (b). This visual comparison highlights the difference in field content between the two scenarios.}\label{fig:difference}
\end{figure}

It is worth noting the fundamental distinction between the negative correction term $\mathcal{I}_c$ in doubly holography and the non-negativity of BMI in a standard CFT, as shown in Fig.~\ref{fig:difference}. In the scenario of a standard CFT -- Fig.~\ref{fig:classical}, one always has the relation: $$\textbf{Area}(\bm{A_1})+\textbf{Area}(\bm{A_2})=\textbf{Area}(\bm{A_1}\cup\bm{A_2}),$$ which implies that in (\ref{eq:mutualinformation}) the quantum fields contributing to $\mathcal{S}[\bm{A_1}] +\mathcal{S}[\bm{A_2}]$ coincide with those entering $ \mathcal{S}[\bm{A_1}\cup \bm{A_2}]$. Since the former generally captures more entanglement, the BMI is manifestly non-negative. In contrast, within the framework of double holography -- Fig.~\ref{fig:semiclassical}, the Q-EWs satisfy the inequality as $$\textbf{Area}(\mathcal{W}_\mathcal{A})+\textbf{Area}(\mathcal{W}_\mathcal{B})\leq\textbf{Area}(\mathcal{W}_{\mathcal{A}\cup\mathcal{B}}).$$ This indicates that the quantum fields contributing to the entropy of  {the individual wedges $\mathcal{W}_\mathcal{A}$ and $\mathcal{W}_\mathcal{B}$} are fewer than those contributing to the combined wedge $\mathcal{W}_{\mathcal{A} \cup \mathcal{B}}$. As a result, the correction term in the BMI, originating from cutoff-independent finite contributions, is generically non-positive. This discrepancy underlines a key qualitative difference between two holographic setups and highlights the nontrivial structure of entanglement in doubly holographic scenarios.

%------------------------------------
%------------------------------------

\section{Interpretation from random tensor networks}\label{sec:RTN}

From the brane perspective in double holography, the negative contribution to the BMI from bulk quantum fields arises from their volume-law entanglement entropy, which itself is a consequence of strong entanglement between the bulk fields and the radiation.

We reproduce this phenomenon using RTNs. As a toy model of holography, a RTN recovers the generalized-entropy formalism in the large bond-dimension limit and realizes a subsystem quantum error-correcting code with complementary recovery \cite{Hayden:2016cfa,Harlow:2016vwg}.

In this section, we first briefly review the RTN construction and its entropy formula. We then show that a maximally mixed state on the bulk degrees of freedom, whose entropy scales with volume, yields a negative contribution to the BMI between boundary subsystems.

\begin{figure}
    \centering
    \includegraphics[width=0.5\linewidth]{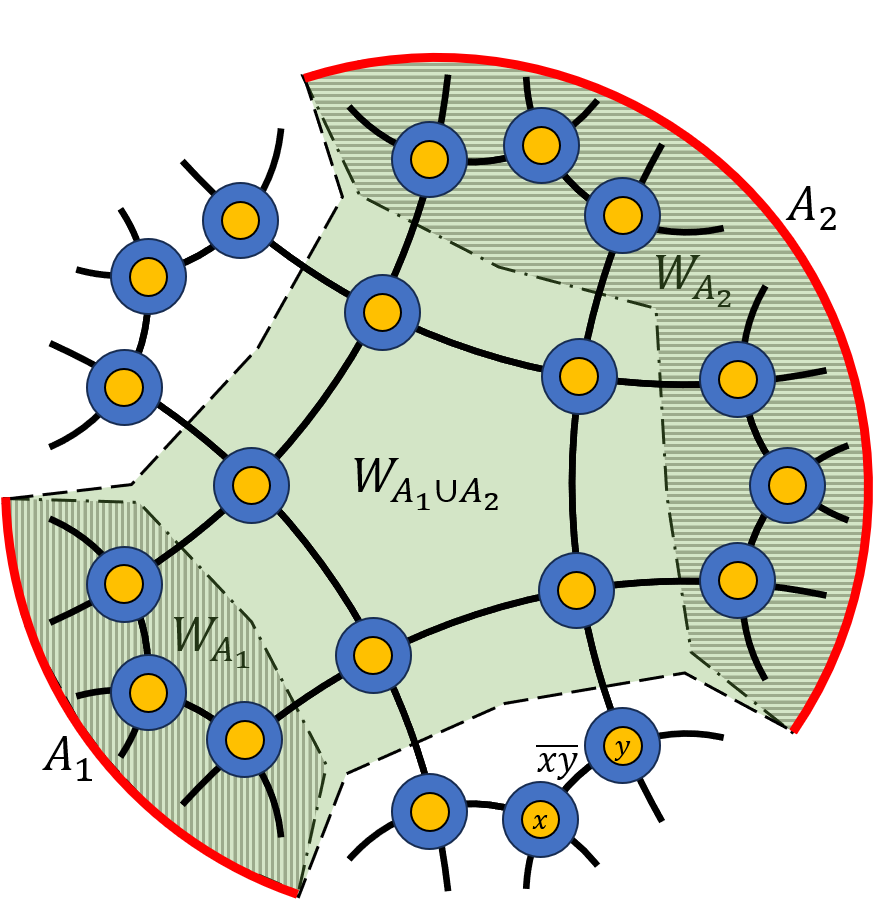}
    \caption{A RTN on the $\ke{4,5}$ tiling of the hyperbolic plane. Each random tensor (blue disk) carries $5$ indices: $4$ of them (black edges) are contracted with other random tensors or extend to the boundary, while $1$ is associated with the local bulk Hilbert space (orange disk). For the boundary region $A_1$ ($A_2$), represented by a red arc, its minimal cut $\gamma_{A_1}$ ($\gamma_{A_2}$) is shown as a dot-dashed line, and its wedge $W_{A_1}$ ($W_{A_2}$) is indicated by the shaded region. For the boundary region $A_1\cup A_2$, the minimal cut $\gamma_{A_1\cup A_2}$ is shown by dashed lines, and its wedge $W_{A_1\cup A_2}$ is indicated by the green region.}
    \label{fig:RTN}
\end{figure}

\subsection{A review of holographic mapping from RTN}

A RTN state is a random projected entangled pair state (PEPS) with boundary degrees of freedom. Consider a network with boundary, parameterized by $\kc{\I,\partial,\E}$, where the network consists of a set of internal vertices $\I$, a set of boundary vertices $\partial$, and a set of edges $\E$ connecting vertices. An internal vertex can be connected to more than one other vertex by edges. A boundary vertex is connected to a single internal vertex by one edge.

Although our construction is general, we mostly consider networks embedded in a two-dimensional manifold with negative curvature, as illustrated in Fig.\,\ref{fig:RTN}, which is the typical case in the context of holography \cite{Hayden:2016cfa}.

We associate a Hilbert space $\H_x$ with each vertex $x\in\I\cup\partial$ and factorize it into Hilbert spaces $\H_{xy}$ associated with its neighboring vertices $y$, together with a local bulk Hilbert space $\H_{xb}$. Explicitly,
\[
\H_x=\kc{\bigotimes_{y:\,\overline{xy}\in \E} \H_{xy}}\otimes \H_{xb},
\]
where $y:\,\overline{xy}\in \E$ denotes a neighboring vertex $y$ connected to $x$ by the edge $\overline{xy}$. The dimensions of these Hilbert spaces are denoted by $d_x=\dim(\H_x)$, $d_{xb}=\dim(\H_{xb})$, and $d_{\overline{xy}}=\dim(\H_{xy})=\dim(\H_{yx})$, where we require $\H_{xy}$ and $\H_{yx}$ to have the same dimension.

From this factorization, we have $d_x=d_{xb}\prod_{y:\overline{xy}\in\E}d_{xy}$. For any boundary vertex $x\in\partial$, we require a trivial local bulk Hilbert space with $d_{xb}=1$, so that $\H_x=\H_{xy}$, where $y\in\I$ is the unique internal vertex connected to $x$. All local bulk Hilbert spaces together form the bulk Hilbert space $\H_\B=\bigotimes_{x\in\I}\H_{xb}$, with total dimension $d_\B=\prod_{x\in\I} d_{xb}$. Similarly, all boundary Hilbert spaces form the boundary Hilbert space $\H_\partial=\bigotimes_{x\in\partial}\H_x$, with dimension $d_\partial=\prod_{x\in\partial} d_x$.

We construct a bulk-to-boundary map based on the RTN. On each internal vertex $x\in\I$, we associate a Haar-random state $\ket{U_x}=U_x\ket{0_x}\in\H_x$, where each $U_x$ is drawn from the circular unitary ensemble (CUE) on $\H_x$, and $\ket{0_x}$ is a fixed reference state. Due to the factorization of $\H_x$, the state $\ket{U_x}$ defines a random tensor with multiple indices associated with the edge Hilbert spaces $\H_{xy}$, and one bulk index associated with $\H_{xb}$.

On each edge $\overline{xy}\in\E$, we place a maximally entangled state
\[
\ket{\overline{xy}}=d_{\overline{xy}}^{-1/2}\sum_{i=1}^{d_{\overline{xy}}} \ket{i}\otimes\ket{i}\in \H_{xy}\otimes\H_{yx},
\]
where $\{\ket{i}\}$ are orthonormal bases of $\H_{xy}$ and $\H_{yx}$. The RTN is then defined as
\[
    V = \kc{\bigotimes_{x\in \I}\sqrt{d_x}\bra{U_x}}
    \kc{\bigotimes_{\overline{xy}\in \E}\ket{\overline{xy}}},
\]
where the factor $\sqrt{d_x}$ is included to ensure the normalization condition \eqref{eq:map_normalization} below. This defines a bulk-to-boundary map $V:\H_\B\to\H_\partial$, such that a bulk state $\rho_\B$ is mapped to a boundary state
\begin{align}\label{eq:bdy_state}
    \rho_\partial=V \rho_\B V^\dagger.
\end{align}

Denoting the ensemble average over the Haar measure as
\[
\overline{X}=\int_{\text{Haar}} X\prod_{x\in\I} dU_x,
\]
with normalization $\overline{1}=1$, we have
\[
    \overline{\ket{U_x}\bra{U_x}}=\frac1{d_x}\mathbb I_x.
\]
It then follows that
\begin{align}\label{eq:map_normalization}
    \overline{V^\dagger V}=\mathbb I_\B,
\end{align}
where $\mathbb I_\B$ is the identity operator on $\H_\B$.

To check whether $V$ is approximately isometric, one may compute the purity \cite{Hayden:2016cfa}
\[
    \frac{\overline{\Tr[(V^\dagger V)^2]}}{\overline{\Tr[V^\dagger V]^2}}
    \approx \frac1{d_\B} + \frac1{d_\partial},\quad     
    d_{xb}\gg1,\ \forall x,\quad d_{\overline{xy}}\gg1,\ \forall\overline{xy}.
\]
When $d_\partial\gg d_\B$, the first term dominates and $V^\dagger V\approx \mathbb I_\B$, so that $V$ becomes an isometric bulk-to-boundary map. Identifying the boundary Hilbert space $\H_\partial$ with the boundary field theory, the bulk Hilbert space $\H_\B$ describes weak fluctuations around a classical geometry, and the isometric map $V$ realizes a one-directional holographic mapping from bulk fluctuations to boundary degrees of freedom \cite{Hayden:2016cfa}\footnote{A boundary-to-bulk isometry can be realized by enlarging the bulk Hilbert space \cite{Yang:2015uoa}.}.

The map $V$ reproduces the Ryu-Takayanagi (RT) formula with bulk-state corrections in the large-dimension limit. For the boundary state \eqref{eq:bdy_state}, the ensemble-averaged $n$-th R\'enyi entropy of a boundary subregion $A$ is
\[
    \S_n[A]=\frac1{1-n}\ln\frac{\overline{\Tr[\rho_A^n]}}{\overline{\Tr[\rho_A]^n}},\quad \rho_A=\Tr_{\bar A}\rho_\partial,
\]
where $\bar A$ denotes the complement of $A$. The replica partition functions can be mapped to a $\mathrm{Sym}_n$ spin model and evaluated in the large-$d_{xb}$ and large-$d_{\overline{xy}}$ limits. For simplicity, we assume uniform dimensions $d_{xb}=d_b$ and $d_{\overline{xy}}=d_e$. The resulting R\'enyi entropy is \cite{Hayden:2016cfa}
\begin{align}\label{eq:entropy_RTN}
    \S_n[A]=\min_{\gamma_A} \kc{\abs{\gamma_A}\ln d_e + S_n[W_A;\rho_\B]},
\end{align}
where $\gamma_A$ is a cut homologous to $A$, enclosing a bulk region $W_A$ with $\partial W_A=A\cup\gamma_A$. This reproduces the quantum extremal surface formula \eqref{eq:EEfromBrane} and provides a RTN interpretation of bulk entropy contributions.

\subsection{Negative bulk contribution to BMI}
To mimic the double-holographic setup,  {we consider a composite quantum system consisting of a large bath $\R$ and a bulk $\B$. The total Hilbert space is thus the tensor product of the two individuals as $\H_\B\otimes\H_\R$. Further specifying the composite system to be in a global vacuum state $\ket{\psi}\in\H_\B\otimes\H_\R$, the bulk and the bath are sufficiently entangled. Consequently, the reduced density matrix on the bulk is given by $\rho_\B=\Tr_\R\ket{\psi}\bra{\psi}$.} When the bath $\R$ is much larger than the bulk system $\B$, the reduced state $\rho_\B$ is expected to be highly mixed. The simplest  {choice} is
\[
    \rho_\B=\frac1{d_\B}\mathbb I_\B.
\]
In this case, the reduced state on any bulk wedge $W_A$ is also maximally mixed, and the R\'enyi entropy obeys a volume law,
\begin{align}\label{eq:volume_law}
    S_n[W_A;\rho_\B]=\abs{W_A}\ln d_b,
\end{align}
where $\abs{W_A}$ denotes the number of vertices in $W_A$.

Following Subsec.\,\ref{sec:EEMI}, we consider two disconnected boundary regions $A_1$ and $A_2$ and compute their R\'enyi BMI in the large-dimension limit. Using \eqref{eq:entropy_RTN}, we find
\begin{align}\label{eq:MI_RTN}
    \I_n[A_1:A_2]&=(\abs{\gamma_{A_1}}+\abs{\gamma_{A_2}}-\abs{\gamma_{A_1\cup A_2}})\ln d_e + \I_{nc}[A_1:A_2], \\
     \I_{nc}[A_1:A_2]&=
     S_n[W_{A_1};\rho_\B]
     +S_n[W_{A_2};\rho_\B]-S_n[W_{A_1\cup A_2};\rho_\B].
\end{align}

Assuming the hierarchy $d_e\gg d_b$, the minimal cuts are determined purely by geometry, and the inequality
\begin{align}\label{eq:gamma_ineq_RTN}
    \abs{\gamma_{A_1}}+\abs{\gamma_{A_2}}\geq \abs{\gamma_{A_1\cup A_2}}
\end{align}
always holds. When $A_1$ and $A_2$ are sufficiently close, the minimal surface $\gamma_{A_1\cup A_2}$ is connected, as shown in Fig.\,\ref{fig:RTN}. In this case, using the volume law \eqref{eq:volume_law}, the bulk contribution to the BMI is non-positive,
\[
    \I_{nc}[A_1:A_2]=(\abs{W_{A_1}}+\abs{W_{A_2}}-\abs{W_{A_1\cup A_2}})\ln d_b\leq 0,
\]
with equality in the disconnected phase.

By contrast, if the bulk entropy obeys an area law,
\[
S_n[W_A;\rho_\B]\sim \abs{\gamma_A}\ln d_e,
\]
then \eqref{eq:gamma_ineq_RTN} implies a non-negative bulk contribution,
\[
    \I_{nc}[A_1:A_2]\sim (\abs{\gamma_{A_1}}+\abs{\gamma_{A_2}}-\abs{\gamma_{A_1\cup A_2}})\ln d_e\geq 0.
\]
We therefore conclude that the non-positive contribution of bulk quantum fields to BMI can be understood, within the RTN framework, as a consequence of the bulk being in a highly mixed, volume-law entangled state.

%------------------------------------
%------------------------------------
\section{Conclusions and discussions}\label{sec:4}
%------------------------------------
%------------------------------------
In this paper, we have investigated the BMI between two one-dimensional spatial subregions $\bm{A_1}$ and $\bm{A_2}$ on the (1+1)-dimensional boundary system $\partial\mathcal{B}$, within the framework of double holography. In this setup, the boundary system $\partial\mathcal{B}$ is coupled to a (2+1)-dimensional heat bath $\partial$. To compute the BMI, we constructed the corresponding RT surface of a disconnected spatial region $\bm{A}=\bm{A_1}\cup\bm{A_2}$, using a shape optimization program, ``Surface Evolver''. This method begins with an initial oriented surface anchored on a spatial slice of AdS$_4$ spacetime with a Planck brane $\mathcal{B}$ and employs a gradient descent algorithm to evolve the surface into an extremal one with minimal area. Notably, Neumann boundary conditions are naturally implemented by constraining the RT surface’s boundary to lie on the brane.

Our analysis is restricted to a specific parameter regime $\theta_0 \leq \theta_c$, where the Q-EW associated with any finite region $\bm{A} \subset \partial\mathcal{B}$ remains finite. For $\theta_0 > \theta_c$, by contrast, the size of the wedge collapses to zero, despite $\bm{A}$ remaining finite \cite{Liu:2023ggg}.

We first validated the convergence and robustness of the numerical method by computing the entanglement entropy of a single spatial region on $\partial\mathcal{B}$. The resulting entropy exhibits both a leading linear divergence and a subleading logarithmic divergence. The leading divergence originates from the brane CFT$_3$ near the conformal boundary, while the logarithmic term arises from the geometric contribution of the area of the QES. The coefficient of this logarithmic term matches the central charge $c'$ of the CFT$_2$ on $\partial\mathcal{B}$, consistent with the results in literature \cite{Liu:2023ggg}. We then extended this analysis to the more technically involved case of a disconnected region $\bm{A} = \bm{A_1} \cup \bm{A_2}$, finding a similar entropy structure in (\ref{eq:entropy2}), again confirming the convergence of the coefficient to $c'$ in the semiclassical limit.

The core of this study has focused on characterizing the entanglement structure between two codimension-three subregions by computing their BMI $\mathcal{I}[\bm{A_1} : \bm{A_2}]$. 
In the semiclassical limit, the BMI can be decomposed into a geometric contribution $\mathcal{I}_g[\bm{A_1} : \bm{A_2}]$ from the areas of the QESs, which is finite and logarithmic, and a correction term $\mathcal{I}_c[\bm{A_1} : \bm{A_2}]$, stemming from quantum fields within the Q-EW. 
Notably, we have found that $\mathcal{I}_g \geq \mathcal{I}$ and $\mathcal{I}c \leq 0$. This negative correction arises because, although the UV components of the quantum fields within $\mathcal{W}_{\bm{A_1}}$, $\mathcal{W}_{\bm{A_2}}$, and $\mathcal{W}_{\bm{A_1} \cup \bm{A_2}}$ are similar, the IR contributions differ significantly: the connected wedge $\mathcal{W}_{\bm{A_1} \cup \bm{A_2}}$ contains obvious more quantum fields than the union of the individual wedges, leading to $\mathcal{I}_c < 0$ as described by Eq.(\ref{eq:CorrectionTerm}) and illustrated in Fig.\ref{fig:semiclassical}.

Earlier work \cite{Liu:2023ggg} conjectured the existence of an intermediate phase, in which the Q-EW on the brane is disconnected but the classical extremal surface in the bulk remains connected. However, our numerical analysis shows that such configurations are unstable: after a finite number of mesh refinements, they invariably converge to one of two stable phases discussed in this work. Therefore, these configurations cannot be regarded as genuine extremal surfaces. The underlying reason for the absence of this intermediate phase deserves further investigation.

Within the RTN framework, the negative contribution of bulk quantum fields to BMI acquires a transparent information-theoretic interpretation. In the large bond-dimension limit, the RTN entropy formula separates naturally into a geometric term determined by minimal cuts and a bulk entropy term associated with degrees of freedom inside Q-EWs. When the bulk is maximally mixed due to strong entanglement with a heat bath, its entropy obeys a volume law. As a consequence, the contribution from bulk fields to BMI becomes non-positive whenever the combined wedge contains strictly more degrees of freedom of fields than the individual wedges.

This analysis clarifies the physical origin of the negative correction term observed in double holography. From the brane perspective, fields on the brane are strongly entangled with the heat bath, leading to highly mixed states inside Q-EWs. The RTN model demonstrates that such volume-law entanglement generically reduces BMI by introducing a negative bulk entropy contribution. By contrast, if the bulk entropy followed an area law, the same RTN construction would yield a non-negative correction, emphasizing that the sign of the bulk term is controlled by the entanglement structure.

Looking forward, it would be of great interest to study other quantum information measures in the context of double holography. In general, any such measure may be decomposed into a geometric component from the induced gravity on the brane and a correction component from the quantum fields on the brane. The geometric contribution can be captured by the QES and reduces to classical behavior in the semiclassical limit. The correction component, involving quantum fields deep into the brane, is generally non-analytic and requires further numerical investigation.

Finally, an important extension would be to explore finite-temperature systems by introducing black holes into the bulk geometry. This involves solving the backreacted brane geometry, possibly via the DeTurck method \cite{Headrick:2009pv, Dias:2015nua}. As is well-known, raising the temperature of the system will disrupt the long-range correlations. Consequently, both the QES and the quantum field contributions would be expected to undergo qualitative changes in their behavior, making this direction particularly compelling for future work.

%------------------------------------
%------------------------------------
\section*{Acknowledgments}
We are grateful to Alexander Jahn and Yuan Sunfor the helpful discussions. Liu Yuxuan special thanks to Peiwen Cao for supporting his work. LYX is supported by the Natural Science Foundation of China under Grant No.~12405079, the Natural Science Foundation of Hunan Province, China (Grant No.~2025JJ60062), and Research start-up funds from the Central South University. YL is supported by the Natural Science Foundation of China under Grant No.~12275275. ZYX also acknowledges support from the berlin Quantum Initiative. 
%------------------------------------
%------------------------------------
\bibliographystyle{unsrt}

\bibliography{ref}

@Misc{evolverlink,
note = {\url{http://www.susqu.edu/brakke/evolver/evolver.html}},
title = {{\it Surface Evolver} program},
author = {Brakke, Kenneth A}
}

@article{brakke1992surface,
  title={The surface evolver},
  author={Brakke, Kenneth A},
  journal={Experimental mathematics},
  volume={1},
  number={2},
  pages={141--165},
  year={1992},
  publisher={Taylor \& Francis}
}

@article{Maldacena:1997re,
    author = "Maldacena, Juan Martin",
    title = "{The Large N limit of superconformal field theories and supergravity}",
    eprint = "hep-th/9711200",
    archivePrefix = "arXiv",
    reportNumber = "HUTP-97-A097, HUTP-98-A097",
    doi = "10.4310/ATMP.1998.v2.n2.a1",
    journal = "Adv. Theor. Math. Phys.",
    volume = "2",
    pages = "231--252",
    year = "1998"
}

@article{Gubser:1998bc,
    author = "Gubser, S. S. and Klebanov, Igor R. and Polyakov, Alexander M.",
    title = "{Gauge theory correlators from noncritical string theory}",
    eprint = "hep-th/9802109",
    archivePrefix = "arXiv",
    reportNumber = "PUPT-1767",
    doi = "10.1016/S0370-2693(98)00377-3",
    journal = "Phys. Lett. B",
    volume = "428",
    pages = "105--114",
    year = "1998"
}

@article{Witten:1998qj,
    author = "Witten, Edward",
    title = "{Anti-de Sitter space and holography}",
    eprint = "hep-th/9802150",
    archivePrefix = "arXiv",
    reportNumber = "IASSNS-HEP-98-15",
    doi = "10.4310/ATMP.1998.v2.n2.a2",
    journal = "Adv. Theor. Math. Phys.",
    volume = "2",
    pages = "253--291",
    year = "1998"
}

@article{Randall:1999vf,
    author = "Randall, Lisa and Sundrum, Raman",
    title = "{An Alternative to compactification}",
    eprint = "hep-th/9906064",
    archivePrefix = "arXiv",
    reportNumber = "MIT-CTP-2874, PUPT-1867, BUHEP-99-13",
    doi = "10.1103/PhysRevLett.83.4690",
    journal = "Phys. Rev. Lett.",
    volume = "83",
    pages = "4690--4693",
    year = "1999"
}

@article{Emparan:1999pm,
    author = "Emparan, Roberto and Johnson, Clifford V. and Myers, Robert C.",
    title = "{Surface terms as counterterms in the AdS / CFT correspondence}",
    eprint = "hep-th/9903238",
    archivePrefix = "arXiv",
    reportNumber = "DTP-99-21, UK-99-04, MCGILL-99-12, EHU-FT-9906",
    doi = "10.1103/PhysRevD.60.104001",
    journal = "Phys. Rev. D",
    volume = "60",
    pages = "104001",
    year = "1999"
}

@article{Gubser:1999vj,
    author = "Gubser, Steven S.",
    title = "{AdS / CFT and gravity}",
    eprint = "hep-th/9912001",
    archivePrefix = "arXiv",
    reportNumber = "HUTP-99-A065",
    doi = "10.1103/PhysRevD.63.084017",
    journal = "Phys. Rev. D",
    volume = "63",
    pages = "084017",
    year = "2001"
}

@article{Dvali:2000hr,
    author = "Dvali, G.R. and Gabadadze, Gregory and Porrati, Massimo",
    title = "{4-D gravity on a brane in 5-D Minkowski space}",
    eprint = "hep-th/0005016",
    archivePrefix = "arXiv",
    reportNumber = "NYU-TH-00-04-01",
    doi = "10.1016/S0370-2693(00)00669-9",
    journal = "Phys. Lett. B",
    volume = "485",
    pages = "208--214",
    year = "2000"
}

@article{Ryu:2006bv,
    author = "Ryu, Shinsei and Takayanagi, Tadashi",
    title = "{Holographic derivation of entanglement entropy from AdS/CFT}",
    eprint = "hep-th/0603001",
    archivePrefix = "arXiv",
    reportNumber = "NSF-KITP-06-11",
    doi = "10.1103/PhysRevLett.96.181602",
    journal = "Phys. Rev. Lett.",
    volume = "96",
    pages = "181602",
    year = "2006"
}

@article{Ryu:2006ef,
    author = "Ryu, Shinsei and Takayanagi, Tadashi",
    title = "{Aspects of Holographic Entanglement Entropy}",
    eprint = "hep-th/0605073",
    archivePrefix = "arXiv",
    reportNumber = "NSF-KITP-06-31, KUNS-2021",
    doi = "10.1088/1126-6708/2006/08/045",
    journal = "JHEP",
    volume = "08",
    pages = "045",
    year = "2006"
}

@article{Hubeny:2007xt,
    author = "Hubeny, Veronika E. and Rangamani, Mukund and Takayanagi, Tadashi",
    title = "{A Covariant holographic entanglement entropy proposal}",
    eprint = "0705.0016",
    archivePrefix = "arXiv",
    primaryClass = "hep-th",
    reportNumber = "DCPT-07-13, KUNS-2069",
    doi = "10.1088/1126-6708/2007/07/062",
    journal = "JHEP",
    volume = "07",
    pages = "062",
    year = "2007"
}

@article{Headrick:2009pv,
    author = "Headrick, Matthew and Kitchen, Sam and Wiseman, Toby",
    title = "{A New approach to static numerical relativity, and its application to Kaluza-Klein black holes}",
    eprint = "0905.1822",
    archivePrefix = "arXiv",
    primaryClass = "gr-qc",
    reportNumber = "BRX-TH-608",
    doi = "10.1088/0264-9381/27/3/035002",
    journal = "Class. Quant. Grav.",
    volume = "27",
    pages = "035002",
    year = "2010"
}

@article{Takayanagi:2011zk,
    author = "Takayanagi, Tadashi",
    title = "{Holographic Dual of BCFT}",
    eprint = "1105.5165",
    archivePrefix = "arXiv",
    primaryClass = "hep-th",
    reportNumber = "IPMU11-0091",
    doi = "10.1103/PhysRevLett.107.101602",
    journal = "Phys. Rev. Lett.",
    volume = "107",
    pages = "101602",
    year = "2011"
}

@article{Omidi:2021opl,
    author = "Omidi, Farzad",
    title = "{Entropy of Hawking radiation for two-sided hyperscaling violating black branes}",
    eprint = "2112.05890",
    archivePrefix = "arXiv",
    primaryClass = "hep-th",
    reportNumber = "IPM/P-2021/41",
    doi = "10.1007/JHEP04(2022)022",
    journal = "JHEP",
    volume = "04",
    pages = "022",
    year = "2022"
}

@article{Lewkowycz:2013nqa,
	author = "Lewkowycz, Aitor and Maldacena, Juan",
	title = "{Generalized gravitational entropy}",
	eprint = "1304.4926",
	archivePrefix = "arXiv",
	primaryClass = "hep-th",
	doi = "10.1007/JHEP08(2013)090",
	journal = "JHEP",
	volume = "08",
	pages = "090",
	year = "2013"
}

@article{Engelhardt:2014gca,
    author = "Engelhardt, Netta and Wall, Aron C.",
    title = "{Quantum Extremal Surfaces: Holographic Entanglement Entropy beyond the Classical Regime}",
    eprint = "1408.3203",
    archivePrefix = "arXiv",
    primaryClass = "hep-th",
    doi = "10.1007/JHEP01(2015)073",
    journal = "JHEP",
    volume = "01",
    pages = "073",
    year = "2015"
}

@article{Fonda:2014cca,
    author = "Fonda, Piermarco and Giomi, Luca and Salvio, Alberto and Tonni, Erik",
    title = "{On shape dependence of holographic mutual information in AdS$_{4}$}",
    eprint = "1411.3608",
    archivePrefix = "arXiv",
    primaryClass = "hep-th",
    reportNumber = "IFT-UAM-CSIC-14-118",
    doi = "10.1007/JHEP02(2015)005",
    journal = "JHEP",
    volume = "02",
    pages = "005",
    year = "2015"
}

@article{Dias:2015nua,
    author = "Dias, Óscar J.C. and Santos, Jorge E. and Way, Benson",
    title = "{Numerical Methods for Finding Stationary Gravitational Solutions}",
    eprint = "1510.02804",
    archivePrefix = "arXiv",
    primaryClass = "hep-th",
    doi = "10.1088/0264-9381/33/13/133001",
    journal = "Class. Quant. Grav.",
    volume = "33",
    number = "13",
    pages = "133001",
    year = "2016"
}

@article{Fonda:2015nma,
    author = "Fonda, Piermarco and Seminara, Domenico and Tonni, Erik",
    title = "{On shape dependence of holographic entanglement entropy in AdS$_{4}$/CFT$_{3}$}",
    eprint = "1510.03664",
    archivePrefix = "arXiv",
    primaryClass = "hep-th",
    doi = "10.1007/JHEP12(2015)037",
    journal = "JHEP",
    volume = "12",
    pages = "037",
    year = "2015"
}

@article{Miao:2017gyt,
    author = "Miao, Rong-Xin and Chu, Chong-Sun and Guo, Wu-Zhong",
    title = "{New proposal for a holographic boundary conformal field theory}",
    eprint = "1701.04275",
    archivePrefix = "arXiv",
    primaryClass = "hep-th",
    reportNumber = "NCTS-TH-1701",
    doi = "10.1103/PhysRevD.96.046005",
    journal = "Phys. Rev. D",
    volume = "96",
    number = "4",
    pages = "046005",
    year = "2017"
}

@article{Chu:2017aab,
    author = "Chu, Chong-Sun and Miao, Rong-Xin and Guo, Wu-Zhong",
    title = "{On New Proposal for Holographic BCFT}",
    eprint = "1701.07202",
    archivePrefix = "arXiv",
    primaryClass = "hep-th",
    reportNumber = "NCTS-TH-1702",
    doi = "10.1007/JHEP04(2017)089",
    journal = "JHEP",
    volume = "04",
    pages = "089",
    year = "2017"
}

@article{Seminara:2017hhh,
    author = "Seminara, Domenico and Sisti, Jacopo and Tonni, Erik",
    title = "{Corner contributions to holographic entanglement entropy in AdS$_{4}$/BCFT$_{3}$}",
    eprint = "1708.05080",
    archivePrefix = "arXiv",
    primaryClass = "hep-th",
    doi = "10.1007/JHEP11(2017)076",
    journal = "JHEP",
    volume = "11",
    pages = "076",
    year = "2017"
}

@article{Seminara:2018pmr,
    author = "Seminara, Domenico and Sisti, Jacopo and Tonni, Erik",
    title = "{Holographic entanglement entropy in AdS$_{4}$/BCFT$_{3}$ and the Willmore functional}",
    eprint = "1805.11551",
    archivePrefix = "arXiv",
    primaryClass = "hep-th",
    doi = "10.1007/JHEP08(2018)164",
    journal = "JHEP",
    volume = "08",
    pages = "164",
    year = "2018"
}

@article{Chu:2018ntx,
    author = "Chu, Chong-Sun and Miao, Rong-Xin",
    title = "{Anomalous Transport in Holographic Boundary Conformal Field Theories}",
    eprint = "1804.01648",
    archivePrefix = "arXiv",
    primaryClass = "hep-th",
    reportNumber = "NCTS-TH-1806, NCTS-TH/1806",
    doi = "10.1007/JHEP07(2018)005",
    journal = "JHEP",
    volume = "07",
    pages = "005",
    year = "2018"
}

@article{Miao:2018qkc,
    author = "Miao, Rong-Xin",
    title = "{Holographic BCFT with Dirichlet Boundary Condition}",
    eprint = "1806.10777",
    archivePrefix = "arXiv",
    primaryClass = "hep-th",
    doi = "10.1007/JHEP02(2019)025",
    journal = "JHEP",
    volume = "02",
    pages = "025",
    year = "2019"
}

@article{Almheiri:2019hni,
	author = "Almheiri, Ahmed and Mahajan, Raghu and Maldacena, Juan and Zhao, Ying",
	title = "{The Page curve of Hawking radiation from semiclassical geometry}",
	eprint = "1908.10996",
	archivePrefix = "arXiv",
	primaryClass = "hep-th",
	doi = "10.1007/JHEP03(2020)149",
	journal = "JHEP",
	volume = "03",
	pages = "149",
	year = "2020"
}

@article{Almheiri:2019psf,
    author = "Almheiri, Ahmed and Engelhardt, Netta and Marolf, Donald and Maxfield, Henry",
    title = "{The entropy of bulk quantum fields and the entanglement wedge of an evaporating black hole}",
    eprint = "1905.08762",
    archivePrefix = "arXiv",
    primaryClass = "hep-th",
    doi = "10.1007/JHEP12(2019)063",
    journal = "JHEP",
    volume = "12",
    pages = "063",
    year = "2019"
}

@article{Chen:2019uhq,
    author = "Chen, Hong Zhe and Fisher, Zachary and Hernandez, Juan and Myers, Robert C. and Ruan, Shan-Ming",
    title = "{Information Flow in Black Hole Evaporation}",
    eprint = "1911.03402",
    archivePrefix = "arXiv",
    primaryClass = "hep-th",
    doi = "10.1007/JHEP03(2020)152",
    journal = "JHEP",
    volume = "03",
    pages = "152",
    year = "2020"
}

@article{Almheiri:2019yqk,
    author = "Almheiri, Ahmed and Mahajan, Raghu and Maldacena, Juan",
    title = "{Islands outside the horizon}",
    journal = "arXiv:1910.11077",
    archivePrefix = "arXiv",
    primaryClass = "hep-th",
    month = "10",
    year = "2019"
}

@article{Penington:2019kki,
    author = "Penington, Geoff and Shenker, Stephen H. and Stanford, Douglas and Yang, Zhenbin",
    title = "{Replica wormholes and the black hole interior}",
    eprint = "1911.11977",
    archivePrefix = "arXiv",
    primaryClass = "hep-th",
    doi = "10.1007/JHEP03(2022)205",
    journal = "JHEP",
    volume = "03",
    pages = "205",
    year = "2022"
}

@article{Almheiri:2019qdq,
    author = "Almheiri, Ahmed and Hartman, Thomas and Maldacena, Juan and Shaghoulian, Edgar and Tajdini, Amirhossein",
    title = "{Replica Wormholes and the Entropy of Hawking Radiation}",
    eprint = "1911.12333",
    archivePrefix = "arXiv",
    primaryClass = "hep-th",
    doi = "10.1007/JHEP05(2020)013",
    journal = "JHEP",
    volume = "05",
    pages = "013",
    year = "2020"
}

@article{Almheiri:2019psy,
    author = "Almheiri, Ahmed and Mahajan, Raghu and Santos, Jorge E.",
    title = "{Entanglement islands in higher dimensions}",
    eprint = "1911.09666",
    archivePrefix = "arXiv",
    primaryClass = "hep-th",
    doi = "10.21468/SciPostPhys.9.1.001",
    journal = "SciPost Phys.",
    volume = "9",
    number = "1",
    pages = "001",
    year = "2020"
}

@article{Chen:2019iro,
    author = "Chen, Yiming",
    title = "{Pulling Out the Island with Modular Flow}",
    eprint = "1912.02210",
    archivePrefix = "arXiv",
    primaryClass = "hep-th",
    doi = "10.1007/JHEP03(2020)033",
    journal = "JHEP",
    volume = "03",
    pages = "033",
    year = "2020"
}

@article{Rozali:2019day,
    author = "Rozali, Moshe and Sully, James and Van Raamsdonk, Mark and Waddell, Christopher and Wakeham, David",
    title = "{Information radiation in BCFT models of black holes}",
    eprint = "1910.12836",
    archivePrefix = "arXiv",
    primaryClass = "hep-th",
    doi = "10.1007/JHEP05(2020)004",
    journal = "JHEP",
    volume = "05",
    pages = "004",
    year = "2020"
}

@article{Cavini:2019wyb,
    author = "Cavini, Giacomo and Seminara, Domenico and Sisti, Jacopo and Tonni, Erik",
    title = "{On shape dependence of holographic entanglement entropy in AdS$_{4}$/CFT$_{3}$ with Lifshitz scaling and hyperscaling violation}",
    eprint = "1907.10030",
    archivePrefix = "arXiv",
    primaryClass = "hep-th",
    doi = "10.1007/JHEP02(2020)172",
    journal = "JHEP",
    volume = "02",
    pages = "172",
    year = "2020"
}

@article{Geng:2020qvw,
    author = "Geng, Hao and Karch, Andreas",
    title = "{Massive islands}",
    eprint = "2006.02438",
    archivePrefix = "arXiv",
    primaryClass = "hep-th",
    doi = "10.1007/JHEP09(2020)121",
    journal = "JHEP",
    volume = "09",
    pages = "121",
    year = "2020"
}

@article{Akal:2020wfl,
    author = "Akal, Ibrahim and Kusuki, Yuya and Takayanagi, Tadashi and Wei, Zixia",
    title = "{Codimension two holography for wedges}",
    eprint = "2007.06800",
    archivePrefix = "arXiv",
    primaryClass = "hep-th",
    reportNumber = "YITP-20-91, IPMU20-0079",
    doi = "10.1103/PhysRevD.102.126007",
    journal = "Phys. Rev. D",
    volume = "102",
    number = "12",
    pages = "126007",
    year = "2020"
}

@article{Miao:2020oey,
    author = "Miao, Rong-Xin",
    title = "{An Exact Construction of Codimension two Holography}",
    eprint = "2009.06263",
    archivePrefix = "arXiv",
    primaryClass = "hep-th",
    doi = "10.1007/JHEP01(2021)150",
    journal = "JHEP",
    volume = "01",
    pages = "150",
    year = "2021"
}

@article{Chen:2020uac,
    author = "Chen, Hong Zhe and Myers, Robert C. and Neuenfeld, Dominik and Reyes, Ignacio A. and Sandor, Joshua",
    title = "{Quantum Extremal Islands Made Easy, Part I: Entanglement on the Brane}",
    eprint = "2006.04851",
    archivePrefix = "arXiv",
    primaryClass = "hep-th",
    doi = "10.1007/JHEP10(2020)166",
    journal = "JHEP",
    volume = "10",
    pages = "166",
    year = "2020"
}

@article{Chen:2020hmv,
    author = "Chen, Hong Zhe and Myers, Robert C. and Neuenfeld, Dominik and Reyes, Ignacio A. and Sandor, Joshua",
    title = "{Quantum Extremal Islands Made Easy, Part II: Black Holes on the Brane}",
    eprint = "2010.00018",
    archivePrefix = "arXiv",
    primaryClass = "hep-th",
    doi = "10.1007/JHEP12(2020)025",
    journal = "JHEP",
    volume = "12",
    pages = "025",
    year = "2020"
}

@article{Hernandez:2020nem,
    author = "Hernandez, Juan and Myers, Robert C. and Ruan, Shan-Ming",
    title = "{Quantum extremal islands made easy. Part III. Complexity on the brane}",
    eprint = "2010.16398",
    archivePrefix = "arXiv",
    primaryClass = "hep-th",
    doi = "10.1007/JHEP02(2021)173",
    journal = "JHEP",
    volume = "02",
    pages = "173",
    year = "2021"
}

@article{Alishahiha:2020qza,
    author = "Alishahiha, Mohsen and Faraji Astaneh, Amin and Naseh, Ali",
    title = "{Island in the presence of higher derivative terms}",
    eprint = "2005.08715",
    archivePrefix = "arXiv",
    primaryClass = "hep-th",
    doi = "10.1007/JHEP02(2021)035",
    journal = "JHEP",
    volume = "02",
    pages = "035",
    year = "2021"
}

@article{Krishnan:2020fer,
    author = "Krishnan, Chethan",
    title = "{Critical Islands}",
    eprint = "2007.06551",
    archivePrefix = "arXiv",
    primaryClass = "hep-th",
    doi = "10.1007/JHEP01(2021)179",
    journal = "JHEP",
    volume = "01",
    pages = "179",
    year = "2021"
}

@article{Balasubramanian:2020hfs,
    author = "Balasubramanian, Vijay and Kar, Arjun and Parrikar, Onkar and S\'arosi, G\'abor and Ugajin, Tomonori",
    title = "{Geometric secret sharing in a model of Hawking radiation}",
    eprint = "2003.05448",
    archivePrefix = "arXiv",
    primaryClass = "hep-th",
    reportNumber = "CERN-TH-2020-039",
    doi = "10.1007/JHEP01(2021)177",
    journal = "JHEP",
    volume = "01",
    pages = "177",
    year = "2021"
}

@article{Hashimoto:2020cas,
    author = "Hashimoto, Koji and Iizuka, Norihiro and Matsuo, Yoshinori",
    title = "{Islands in Schwarzschild black holes}",
    eprint = "2004.05863",
    archivePrefix = "arXiv",
    primaryClass = "hep-th",
    reportNumber = "OU-HET-1053",
    doi = "10.1007/JHEP06(2020)085",
    journal = "JHEP",
    volume = "06",
    pages = "085",
    year = "2020"
}

@article{Almheiri:2020cfm,
    author = "Almheiri, Ahmed and Hartman, Thomas and Maldacena, Juan and Shaghoulian, Edgar and Tajdini, Amirhossein",
    title = "{The entropy of Hawking radiation}",
    eprint = "2006.06872",
    archivePrefix = "arXiv",
    primaryClass = "hep-th",
    doi = "10.1103/RevModPhys.93.035002",
    journal = "Rev. Mod. Phys.",
    volume = "93",
    number = "3",
    pages = "035002",
    year = "2021"
}

@article{Geng:2020fxl,
    author = "Geng, Hao and Karch, Andreas and Perez-Pardavila, Carlos and Raju, Suvrat and Randall, Lisa and Riojas, Marcos and Shashi, Sanjit",
    title = "{Information Transfer with a Gravitating Bath}",
    eprint = "2012.04671",
    archivePrefix = "arXiv",
    primaryClass = "hep-th",
    doi = "10.21468/SciPostPhys.10.5.103",
    journal = "SciPost Phys.",
    volume = "10",
    number = "5",
    pages = "103",
    year = "2021"
}

@article{Ling:2020laa,
    author = "Ling, Yi and Liu, Yuxuan and Xian, Zhuo-Yu",
    title = "{Island in Charged Black Holes}",
    eprint = "2010.00037",
    archivePrefix = "arXiv",
    primaryClass = "hep-th",
    doi = "10.1007/JHEP03(2021)251",
    journal = "JHEP",
    volume = "03",
    pages = "251",
    year = "2021"
}

@article{Anegawa:2020ezn,
    author = "Anegawa, Takanori and Iizuka, Norihiro",
    title = "{Notes on islands in asymptotically flat 2d dilaton black holes}",
    eprint = "2004.01601",
    archivePrefix = "arXiv",
    primaryClass = "hep-th",
    reportNumber = "OU-HET-1051",
    doi = "10.1007/JHEP07(2020)036",
    journal = "JHEP",
    volume = "07",
    pages = "036",
    year = "2020"
}

@article{Hartman:2020swn,
    author = "Hartman, Thomas and Shaghoulian, Edgar and Strominger, Andrew",
    title = "{Islands in Asymptotically Flat 2D Gravity}",
    eprint = "2004.13857",
    archivePrefix = "arXiv",
    primaryClass = "hep-th",
    doi = "10.1007/JHEP07(2020)022",
    journal = "JHEP",
    volume = "07",
    pages = "022",
    year = "2020"
}

@article{Caceres:2020jcn,
    author = "Caceres, Elena and Kundu, Arnab and Patra, Ayan K. and Shashi, Sanjit",
    title = "{Warped information and entanglement islands in AdS/WCFT}",
    eprint = "2012.05425",
    archivePrefix = "arXiv",
    primaryClass = "hep-th",
    reportNumber = "UTTG-23-2020",
    doi = "10.1007/JHEP07(2021)004",
    journal = "JHEP",
    volume = "07",
    pages = "004",
    year = "2021"
}

@article{Li:2020ceg,
    author = "Li, Tianyi and Chu, Jinwei and Zhou, Yang",
    title = "{Reflected Entropy for an Evaporating Black Hole}",
    eprint = "2006.10846",
    archivePrefix = "arXiv",
    primaryClass = "hep-th",
    doi = "10.1007/JHEP11(2020)155",
    journal = "JHEP",
    volume = "11",
    pages = "155",
    year = "2020"
}

@article{Gautason:2020tmk,
    author = "Gautason, F. F. and Schneiderbauer, Lukas and Sybesma, Watse and Thorlacius, L\'arus",
    title = "{Page Curve for an Evaporating Black Hole}",
    eprint = "2004.00598",
    archivePrefix = "arXiv",
    primaryClass = "hep-th",
    doi = "10.1007/JHEP05(2020)091",
    journal = "JHEP",
    volume = "05",
    pages = "091",
    year = "2020"
}

@article{Krishnan:2020oun,
    author = "Krishnan, Chethan and Patil, Vaishnavi and Pereira, Jude",
    title = "{Page Curve and the Information Paradox in Flat Space}",
    eprint = "2005.02993",
    archivePrefix = "arXiv",
    primaryClass = "hep-th",
    month = "5",
    year = "2020"
}

@article{Karlsson:2020uga,
    author = "Karlsson, Anna",
    title = "{Replica wormhole and island incompatibility with monogamy of entanglement}",
    eprint = "2007.10523",
    archivePrefix = "arXiv",
    primaryClass = "hep-th",
    month = "7",
    year = "2020"
}

@article{Marolf:2020xie,
    author = "Marolf, Donald and Maxfield, Henry",
    title = "{Transcending the ensemble: baby universes, spacetime wormholes, and the order and disorder of black hole information}",
    eprint = "2002.08950",
    archivePrefix = "arXiv",
    primaryClass = "hep-th",
    doi = "10.1007/JHEP08(2020)044",
    journal = "JHEP",
    volume = "08",
    pages = "044",
    year = "2020"
}

@article{Balasubramanian:2020jhl,
    author = "Balasubramanian, Vijay and Kar, Arjun and Ross, Simon F. and Ugajin, Tomonori",
    title = "{Spin structures and baby universes}",
    eprint = "2007.04333",
    archivePrefix = "arXiv",
    primaryClass = "hep-th",
    doi = "10.1007/JHEP09(2020)192",
    journal = "JHEP",
    volume = "09",
    pages = "192",
    year = "2020"
}

@article{Chen:2020jvn,
    author = "Chen, Hong Zhe and Fisher, Zachary and Hernandez, Juan and Myers, Robert C. and Ruan, Shan-Ming",
    title = "{Evaporating Black Holes Coupled to a Thermal Bath}",
    eprint = "2007.11658",
    archivePrefix = "arXiv",
    primaryClass = "hep-th",
    doi = "10.1007/JHEP01(2021)065",
    journal = "JHEP",
    volume = "01",
    pages = "065",
    year = "2021"
}

@article{Balasubramanian:2020xqf,
    author = "Balasubramanian, Vijay and Kar, Arjun and Ugajin, Tomonori",
    title = "{Islands in de Sitter space}",
    eprint = "2008.05275",
    archivePrefix = "arXiv",
    primaryClass = "hep-th",
    doi = "10.1007/JHEP02(2021)072",
    journal = "JHEP",
    volume = "02",
    pages = "072",
    year = "2021"
}

@article{Bhattacharya:2020uun,
    author = "Bhattacharya, Aranya and Chanda, Anindya and Maulik, Sabyasachi and Northe, Christian and Roy, Shibaji",
    title = "{Topological shadows and complexity of islands in multiboundary wormholes}",
    eprint = "2010.04134",
    archivePrefix = "arXiv",
    primaryClass = "hep-th",
    doi = "10.1007/JHEP02(2021)152",
    journal = "JHEP",
    volume = "02",
    pages = "152",
    year = "2021"
}

@article{Deng:2020ent,
    author = "Deng, Feiyu and Chu, Jinwei and Zhou, Yang",
    title = "{Defect extremal surface as the holographic counterpart of Island formula}",
    eprint = "2012.07612",
    archivePrefix = "arXiv",
    primaryClass = "hep-th",
    doi = "10.1007/JHEP03(2021)008",
    journal = "JHEP",
    volume = "03",
    pages = "008",
    year = "2021"
}

@article{Akal:2020twv,
    author = "Akal, Ibrahim and Kusuki, Yuya and Shiba, Noburo and Takayanagi, Tadashi and Wei, Zixia",
    title = "{Entanglement Entropy in a Holographic Moving Mirror and the Page Curve}",
    eprint = "2011.12005",
    archivePrefix = "arXiv",
    primaryClass = "hep-th",
    reportNumber = "YITP-20-149, IPMU20-0121",
    doi = "10.1103/PhysRevLett.126.061604",
    journal = "Phys. Rev. Lett.",
    volume = "126",
    number = "6",
    pages = "061604",
    year = "2021"
}

@article{Sybesma:2020fxg,
    author = "Sybesma, Watse",
    title = "{Pure de Sitter space and the island moving back in time}",
    eprint = "2008.07994",
    archivePrefix = "arXiv",
    primaryClass = "hep-th",
    doi = "10.1088/1361-6382/abff9a",
    journal = "Class. Quant. Grav.",
    volume = "38",
    number = "14",
    pages = "145012",
    year = "2021"
}

@article{Marolf:2020rpm,
    author = "Marolf, Donald and Maxfield, Henry",
    title = "{Observations of Hawking radiation: the Page curve and baby universes}",
    eprint = "2010.06602",
    archivePrefix = "arXiv",
    primaryClass = "hep-th",
    doi = "10.1007/JHEP04(2021)272",
    journal = "JHEP",
    volume = "04",
    pages = "272",
    year = "2021"
}

@article{Anderson:2020vwi,
    author = "Anderson, Louise and Parrikar, Onkar and Soni, Ronak M.",
    title = "{Islands with gravitating baths: towards ER = EPR}",
    eprint = "2103.14746",
    archivePrefix = "arXiv",
    primaryClass = "hep-th",
    doi = "10.1007/JHEP10(2021)226",
    journal = "JHEP",
    volume = "21",
    pages = "226",
    year = "2020"
}

@article{Balasubramanian:2020coy,
    author = "Balasubramanian, Vijay and Kar, Arjun and Ugajin, Tomonori",
    title = "{Entanglement between two disjoint universes}",
    eprint = "2008.05274",
    archivePrefix = "arXiv",
    primaryClass = "hep-th",
    doi = "10.1007/JHEP02(2021)136",
    journal = "JHEP",
    volume = "02",
    pages = "136",
    year = "2021"
}

@article{KumarBasak:2020ams,
    author = "Kumar Basak, Jaydeep and Basu, Debarshi and Malvimat, Vinay and Parihar, Himanshu and Sengupta, Gautam",
    title = "{Islands for entanglement negativity}",
    eprint = "2012.03983",
    archivePrefix = "arXiv",
    primaryClass = "hep-th",
    doi = "10.21468/SciPostPhys.12.1.003",
    journal = "SciPost Phys.",
    volume = "12",
    number = "1",
    pages = "003",
    year = "2022"
}

@article{Ling:2021vxe,
    author = "Ling, Yi and Liu, Peng and Liu, Yuxuan and Niu, Chao and Xian, Zhuo-Yu and Zhang, Cheng-Yong",
    title = "{Reflected entropy in double holography}",
    eprint = "2109.09243",
    archivePrefix = "arXiv",
    primaryClass = "hep-th",
    doi = "10.1007/JHEP02(2022)037",
    journal = "JHEP",
    volume = "02",
    pages = "037",
    year = "2022"
}

@article{Wang:2021woy,
    author = "Wang, Xuanhua and Li, Ran and Wang, Jin",
    title = {{Islands and Page curves of Reissner-Nordstr\"om black holes}},
    eprint = "2101.06867",
    archivePrefix = "arXiv",
    primaryClass = "hep-th",
    doi = "10.1007/JHEP04(2021)103",
    journal = "JHEP",
    volume = "04",
    pages = "103",
    year = "2021"
}

@article{He:2021mst,
    author = "He, Song and Sun, Yuan and Zhao, Long and Zhang, Yu-Xuan",
    title = "{The universality of islands outside the horizon}",
    eprint = "2110.07598",
    archivePrefix = "arXiv",
    primaryClass = "hep-th",
    doi = "10.1007/JHEP05(2022)047",
    journal = "JHEP",
    volume = "05",
    pages = "047",
    year = "2022"
}

@article{Hollowood:2021lsw,
    author = "Hollowood, Timothy J. and Kumar, S. Prem and Legramandi, Andrea and Talwar, Neil",
    title = "{Grey-body factors, irreversibility and multiple island saddles}",
    eprint = "2111.02248",
    archivePrefix = "arXiv",
    primaryClass = "hep-th",
    doi = "10.1007/JHEP03(2022)110",
    journal = "JHEP",
    volume = "03",
    pages = "110",
    year = "2022"
}

@article{Vardhan:2021mdy,
    author = "Vardhan, Shreya and Kudler-Flam, Jonah and Shapourian, Hassan and Liu, Hong",
    title = "{Mixed-state entanglement and information recovery in thermalized states and evaporating black holes}",
    eprint = "2112.00020",
    archivePrefix = "arXiv",
    primaryClass = "hep-th",
    reportNumber = "MIT-CTP/5362",
    doi = "10.1007/JHEP01(2023)064",
    journal = "JHEP",
    volume = "01",
    pages = "064",
    year = "2023"
}

@article{Kawabata:2021vyo,
    author = "Kawabata, Kohki and Nishioka, Tatsuma and Okuyama, Yoshitaka and Watanabe, Kento",
    title = "{Replica wormholes and capacity of entanglement}",
    eprint = "2105.08396",
    archivePrefix = "arXiv",
    primaryClass = "hep-th",
    reportNumber = "YITP-21-45",
    doi = "10.1007/JHEP10(2021)227",
    journal = "JHEP",
    volume = "10",
    pages = "227",
    year = "2021"
}

@article{Kawabata:2021hac,
    author = "Kawabata, Kohki and Nishioka, Tatsuma and Okuyama, Yoshitaka and Watanabe, Kento",
    title = "{Probing Hawking radiation through capacity of entanglement}",
    eprint = "2102.02425",
    archivePrefix = "arXiv",
    primaryClass = "hep-th",
    reportNumber = "YITP-21-08",
    doi = "10.1007/JHEP05(2021)062",
    journal = "JHEP",
    volume = "05",
    pages = "062",
    year = "2021"
}

@article{Geng:2021iyq,
    author = {Geng, Hao and L\"ust, Severin and Mishra, Rashmish K. and Wakeham, David},
    title = "{Holographic BCFTs and Communicating Black Holes}",
    eprint = "2104.07039",
    archivePrefix = "arXiv",
    primaryClass = "hep-th",
    doi = "10.1007/JHEP08(2021)003",
    journal = "jhep",
    volume = "08",
    pages = "003",
    year = "2021"
}

@article{Geng:2021mic,
    author = "Geng, Hao and Karch, Andreas and Perez-Pardavila, Carlos and Raju, Suvrat and Randall, Lisa and Riojas, Marcos and Shashi, Sanjit",
    title = "{Entanglement phase structure of a holographic BCFT in a black hole background}",
    eprint = "2112.09132",
    archivePrefix = "arXiv",
    primaryClass = "hep-th",
    reportNumber = "UTTG-27-2021",
    doi = "10.1007/JHEP05(2022)153",
    journal = "JHEP",
    volume = "05",
    pages = "153",
    year = "2022"
}

@article{Chou:2021boq,
    author = "Chou, Chia-Jui and Lao, Hans B. and Yang, Yi",
    title = "{Page curve of effective Hawking radiation}",
    eprint = "2111.14551",
    archivePrefix = "arXiv",
    primaryClass = "hep-th",
    doi = "10.1103/PhysRevD.106.066008",
    journal = "Phys. Rev. D",
    volume = "106",
    number = "6",
    pages = "066008",
    year = "2022"
}

@article{Chou:2023adi,
    author = "Chou, Chia-Jui and Lao, Hans B. and Yang, Yi",
    title = "{Page Curve of AdS-Vaidya Model for Evaporating Black Holes}",
    eprint = "2306.16744",
    archivePrefix = "arXiv",
    primaryClass = "hep-th",
    month = "6",
    year = "2023"
}

@article{Miyata:2021qsm,
    author = "Miyata, Akihiro and Ugajin, Tomonori",
    title = "{Entanglement between two evaporating black holes}",
    eprint = "2111.11688",
    archivePrefix = "arXiv",
    primaryClass = "hep-th",
    reportNumber = "UT-Komaba/21-5",
    doi = "10.1007/JHEP09(2022)009",
    journal = "JHEP",
    volume = "09",
    pages = "009",
    year = "2022"
}

@article{Akal:2021dqt,
    author = "Akal, Ibrahim and Kawamoto, Taishi and Ruan, Shan-Ming and Takayanagi, Tadashi and Wei, Zixia",
    title = "{Page curve under final state projection}",
    eprint = "2112.08433",
    archivePrefix = "arXiv",
    primaryClass = "hep-th",
    reportNumber = "YITP-21-158, IPMU21-0086, YITP-21-158; IPMU21-0086",
    doi = "10.1103/PhysRevD.105.126026",
    journal = "Phys. Rev. D",
    volume = "105",
    number = "12",
    pages = "126026",
    year = "2022"
}

@article{Renner:2021qbe,
    author = "Renner, Renato and Wang, Jinzhao",
    title = "{The black hole information puzzle and the quantum de Finetti theorem}",
    eprint = "2110.14653",
    archivePrefix = "arXiv",
    primaryClass = "hep-th",
    month = "10",
    year = "2021"
}

@article{Balasubramanian:2021wgd,
    author = "Balasubramanian, Vijay and Kar, Arjun and Ugajin, Tomonori",
    title = "{Entanglement between two gravitating universes}",
    eprint = "2104.13383",
    archivePrefix = "arXiv",
    primaryClass = "hep-th",
    reportNumber = "YITP-21-39",
    doi = "10.1088/1361-6382/ac3c8b",
    journal = "Class. Quant. Grav.",
    volume = "39",
    number = "17",
    pages = "174001",
    year = "2022"
}

@article{Bhattacharya:2021nqj,
    author = "Bhattacharya, Aranya and Bhattacharyya, Arpan and Nandy, Pratik and Patra, Ayan K.",
    title = "{Bath deformations, islands, and holographic complexity}",
    eprint = "2112.06967",
    archivePrefix = "arXiv",
    primaryClass = "hep-th",
    doi = "10.1103/PhysRevD.105.066019",
    journal = "Phys. Rev. D",
    volume = "105",
    number = "6",
    pages = "066019",
    year = "2022"
}

@article{Bhattacharya:2021dnd,
    author = "Bhattacharya, Aranya and Bhattacharyya, Arpan and Nandy, Pratik and Patra, Ayan K.",
    title = "{Partial islands and subregion complexity in geometric secret-sharing model}",
    eprint = "2109.07842",
    archivePrefix = "arXiv",
    primaryClass = "hep-th",
    doi = "10.1007/JHEP12(2021)091",
    journal = "JHEP",
    volume = "12",
    pages = "091",
    year = "2021"
}

@article{Caceres:2021fuw,
    author = "Caceres, Elena and Kundu, Arnab and Patra, Ayan K. and Shashi, Sanjit",
    title = "{Page curves and bath deformations}",
    eprint = "2107.00022",
    archivePrefix = "arXiv",
    primaryClass = "hep-th",
    doi = "10.21468/SciPostPhysCore.5.2.033",
    journal = "SciPost Phys. Core",
    volume = "5",
    pages = "033",
    year = "2022"
}

@article{Bhattacharya:2021jrn,
    author = "Bhattacharya, Aranya and Bhattacharyya, Arpan and Nandy, Pratik and Patra, Ayan K.",
    title = "{Islands and complexity of eternal black hole and radiation subsystems for a doubly holographic model}",
    eprint = "2103.15852",
    archivePrefix = "arXiv",
    primaryClass = "hep-th",
    doi = "10.1007/JHEP05(2021)135",
    journal = "JHEP",
    volume = "05",
    pages = "135",
    year = "2021"
}

@article{Peng:2021vhs,
    author = "Peng, Cheng and Tian, Jia and Yu, Jianghui",
    title = "{Baby universes, ensemble averages and factorizations with matters}",
    eprint = "2111.14856",
    archivePrefix = "arXiv",
    primaryClass = "hep-th",
    journal={arXiv preprint arXiv:2111.14856},
    month = "11",
    year = "2021"
}

@article{Peng:2022pfa,
    author = "Peng, Cheng and Tian, Jia and Yang, Yingyu",
    title = "{Half-wormholes and ensemble averages}",
    eprint = "2205.01288",
    archivePrefix = "arXiv",
    primaryClass = "hep-th",
    doi = "10.1140/epjc/s10052-023-12164-9",
    journal = "Eur. Phys. J. C",
    volume = "83",
    number = "11",
    pages = "993",
    year = "2023"
}

@article{Miyata:2021ncm,
    author = "Miyata, Akihiro and Ugajin, Tomonori",
    title = "{Evaporation of black holes in flat space entangled with an auxiliary universe}",
    eprint = "2104.00183",
    archivePrefix = "arXiv",
    primaryClass = "hep-th",
    reportNumber = "UT-Komaba/21-3, YITP-21-27",
    doi = "10.1093/ptep/ptab163",
    journal = "PTEP",
    volume = "2022",
    number = "1",
    pages = "013B13",
    year = "2022"
}

@article{Grimaldi:2022suv,
    author = "Grimaldi, Guglielmo and Hernandez, Juan and Myers, Robert C.",
    title = "{Quantum extremal islands made easy. Part IV. Massive black holes on the brane}",
    eprint = "2202.00679",
    archivePrefix = "arXiv",
    primaryClass = "hep-th",
    reportNumber = "BRX-TH-6696",
    doi = "10.1007/JHEP03(2022)136",
    journal = "JHEP",
    volume = "03",
    pages = "136",
    year = "2022"
}

@article{Engelhardt:2022qts,
    author = "Engelhardt, Netta and Folkestad, \r{A}smund",
    title = "{Canonical purification of evaporating black holes}",
    eprint = "2201.08395",
    archivePrefix = "arXiv",
    primaryClass = "hep-th",
    reportNumber = "MIT-CTP/5394",
    doi = "10.1103/PhysRevD.105.086010",
    journal = "Phys. Rev. D",
    volume = "105",
    number = "8",
    pages = "086010",
    year = "2022"
}

@article{Suzuki:2022xwv,
    author = "Suzuki, Kenta and Takayanagi, Tadashi",
    title = "{BCFT and Islands in two dimensions}",
    eprint = "2202.08462",
    archivePrefix = "arXiv",
    primaryClass = "hep-th",
    reportNumber = "YITP-22-14, IPMU22-0002",
    doi = "10.1007/JHEP06(2022)095",
    journal = "JHEP",
    volume = "06",
    pages = "095",
    year = "2022"
}

@article{Suzuki:2022yru,
    author = "Suzuki, Yu-ki and Terashima, Seiji",
    title = "{On the dynamics in the AdS/BCFT correspondence}",
    eprint = "2205.10600",
    archivePrefix = "arXiv",
    primaryClass = "hep-th",
    reportNumber = "YITP-22-50",
    doi = "10.1007/JHEP09(2022)103",
    journal = "JHEP",
    volume = "09",
    pages = "103",
    year = "2022"
}

@article{Afrasiar:2022ebi,
    author = "Afrasiar, Mir and Kumar Basak, Jaydeep and Chandra, Ashish and Sengupta, Gautam",
    title = "{Islands for Entanglement Negativity in Communicating Black Holes}",
    journal = "arXiv:2205.07903",
    archivePrefix = "arXiv",
    primaryClass = "hep-th",
    month = "5",
    year = "2022"
}

@article{Izumi:2022opi,
    author = "Izumi, Keisuke and Shiromizu, Tetsuya and Suzuki, Kenta and Takayanagi, Tadashi and Tanahashi, Norihiro",
    title = "{Brane dynamics of holographic BCFTs}",
    eprint = "2205.15500",
    archivePrefix = "arXiv",
    primaryClass = "hep-th",
    reportNumber = "YITP-22-56, IPMU22-0032",
    doi = "10.1007/JHEP10(2022)050",
    journal = "JHEP",
    volume = "10",
    pages = "050",
    year = "2022"
}

@article{Liu:2022pan,
    author = "Liu, Yuxuan and Xian, Zhuo-Yu and Peng, Cheng and Ling, Yi",
    title = "{Addendum to: Black holes entangled by radiation}",
    eprint = "2205.14596",
    archivePrefix = "arXiv",
    primaryClass = "hep-th",
    doi = "10.1007/JHEP11(2022)043",
    journal = "JHEP",
    volume = "11",
    pages = "043",
    year = "2022"
}

@article{Witten:2021unn,
    author = "Witten, Edward",
    title = "{Gravity and the crossed product}",
    eprint = "2112.12828",
    archivePrefix = "arXiv",
    primaryClass = "hep-th",
    doi = "10.1007/JHEP10(2022)008",
    journal = "JHEP",
    volume = "10",
    pages = "008",
    year = "2022"
}

@article{Erdmenger:2014xya,
    author = "Erdmenger, Johanna and Flory, Mario and Newrzella, Max-Niklas",
    title = "{Bending branes for DCFT in two dimensions}",
    eprint = "1410.7811",
    archivePrefix = "arXiv",
    primaryClass = "hep-th",
    reportNumber = "MPP-2014-372",
    doi = "10.1007/JHEP01(2015)058",
    journal = "JHEP",
    volume = "01",
    pages = "058",
    year = "2015"
}

@article{Erdmenger:2015spo,
    author = "Erdmenger, Johanna and Flory, Mario and Hoyos, Carlos and Newrzella, Max-Niklas and Wu, Jackson M. S.",
    title = "{Entanglement Entropy in a Holographic Kondo Model}",
    eprint = "1511.03666",
    archivePrefix = "arXiv",
    primaryClass = "hep-th",
    reportNumber = "MPP-2015-248, FPAUO-15-15",
    doi = "10.1002/prop.201500099",
    journal = "Fortsch. Phys.",
    volume = "64",
    pages = "109--130",
    year = "2016"
}

@article{Jeong:2023lkc,
    author = "Jeong, Hyun-Sik and Kim, Keun-Young and Sun, Ya-Wen",
    title = "{Entanglement entropy analysis of dyonic black holes using doubly holographic theory}",
    eprint = "2305.18122",
    archivePrefix = "arXiv",
    primaryClass = "hep-th",
    reportNumber = "IFT-UAM/CSIC-23-54",
    doi = "10.1103/PhysRevD.108.126016",
    journal = "Phys. Rev. D",
    volume = "108",
    number = "12",
    pages = "126016",
    year = "2023"
}

@article{Ahn:2021chg,
    author = "Ahn, Byoungjoon and Bak, Sang-Eon and Jeong, Hyun-Sik and Kim, Keun-Young and Sun, Ya-Wen",
    title = "{Islands in charged linear dilaton black holes}",
    eprint = "2107.07444",
    archivePrefix = "arXiv",
    primaryClass = "hep-th",
    doi = "10.1103/PhysRevD.105.046012",
    journal = "Phys. Rev. D",
    volume = "105",
    number = "4",
    pages = "046012",
    year = "2022"
}

@article{Uhlemann:2021nhu,
    author = "Uhlemann, Christoph F.",
    title = "{Islands and Page curves in 4d from Type IIB}",
    eprint = "2105.00008",
    archivePrefix = "arXiv",
    primaryClass = "hep-th",
    reportNumber = "LCTP-21-09",
    doi = "10.1007/JHEP08(2021)104",
    journal = "JHEP",
    volume = "08",
    pages = "104",
    year = "2021"
}

@article{Karch:2022rvr,
    author = "Karch, Andreas and Sun, Haoyu and Uhlemann, Christoph F.",
    title = "{Double holography in string theory}",
    eprint = "2206.11292",
    archivePrefix = "arXiv",
    primaryClass = "hep-th",
    reportNumber = "LCTP-22-08",
    doi = "10.1007/JHEP10(2022)012",
    journal = "JHEP",
    volume = "10",
    pages = "012",
    year = "2022"
}

@article{Liu:2024cmv,
    author = "Liu, Yuxuan and Jian, Shao-Kai and Ling, Yi and Xian, Zhuo-Yu",
    title = "{Entanglement inside a black hole before the Page time}",
    eprint = "2401.04706",
    archivePrefix = "arXiv",
    primaryClass = "hep-th",
    month = "1",
    year = "2024"
}

@article{Liu:2023ggg,
    author = "Liu, Yuxuan and Chen, Qian and Ling, Yi and Peng, Cheng and Tian, Yu and Xian, Zhuo-Yu",
    title = "{Addendum to: Entanglement of defect subregions in double holography}",
    eprint = "2312.08025",
    archivePrefix = "arXiv",
    primaryClass = "hep-th",
    doi = "10.1007/JHEP09(2024)194",
    journal = "JHEP",
    volume = "09",
    pages = "194",
    year = "2024"
}

@article{Hayden:2016cfa,
    author = "Hayden, Patrick and Nezami, Sepehr and Qi, Xiao-Liang and Thomas, Nathaniel and Walter, Michael and Yang, Zhao",
    title = "{Holographic duality from random tensor networks}",
    eprint = "1601.01694",
    archivePrefix = "arXiv",
    primaryClass = "hep-th",
    doi = "10.1007/JHEP11(2016)009",
    journal = "JHEP",
    volume = "11",
    pages = "009",
    year = "2016"
}

@article{Harlow:2016vwg,
    author = "Harlow, Daniel",
    title = "{The Ryu{\textendash}Takayanagi Formula from Quantum Error Correction}",
    eprint = "1607.03901",
    archivePrefix = "arXiv",
    primaryClass = "hep-th",
    doi = "10.1007/s00220-017-2904-z",
    journal = "Commun. Math. Phys.",
    volume = "354",
    number = "3",
    pages = "865--912",
    year = "2017"
}

@article{Yang:2015uoa,
    author = "Yang, Zhao and Hayden, Patrick and Qi, Xiao-Liang",
    title = "{Bidirectional holographic codes and sub-AdS locality}",
    eprint = "1510.03784",
    archivePrefix = "arXiv",
    primaryClass = "hep-th",
    doi = "10.1007/JHEP01(2016)175",
    journal = "JHEP",
    volume = "01",
    pages = "175",
    year = "2016"
}

\end{document}